\documentclass[a4paper,fleqn,usenatbib]{mnras}

\usepackage{newtxtext,newtxmath}
\usepackage[T1]{fontenc}
\usepackage{ae,aecompl}
\usepackage{todonotes}

\usepackage{graphicx}	% Including figure files
\usepackage{amsmath}	% Advanced maths commands
\usepackage{amssymb}	% Extra maths symbols

\title[Historical Novae, Classical Novae, and Far Eastern guest stars]{Which information can we derive from historical Far Eastern guest stars for modern research on novae and cataclysmic variables?}

\author[Hoffmann, S.]{
Susanne M. Hoffmann,$^{1}$\thanks{E-mail: susanne.hoffmann@uni-jena.de (SMH)}
\\
$^{1}$Friedrich Schiller University, Faculty of Physics and Astronomy, Max-Wien-Platz 1, Jena, Germany\\
}

\date{Accepted 2019 October 07. Received 2019 October 07; in original form 2019 March 27}

\pubyear{2019}

\begin{document}
\label{firstpage}
\pagerange{\pageref{firstpage}--\pageref{lastpage}}
\maketitle

\begin{abstract}
 Recently, there have been several studies on the evolution of binary systems using historical data treated as facts in the chain of arguments. In this paper we discuss six case studies of modern dwarf novae with suggested historical counterpart from the historical point of view as well as the derived consequences for the physics of close binary systems (the dwarf novae Z Cam and AT Cnc, the nebula in M22, and the possible Nova\,101, Nova\,483, and Nova\,1437). We consider the historical Far Eastern reports and after a careful re-reading of the text we map the given information on the sky. In some cases, the positions given in modern lists of classical nova-guest star-pairs turn out to be wrong, or they have to be considered highly approximate: The historical position, in most cases, should be transformed into areas at the celestial sphere and not into point coordinates. Based on the correct information we consider the consequences concerning the evolution of close binary systems. However, the result is that none of the cases of cataclysmic variables suggested to have a historical counterpart can be (fully) supported. As the identification of the historical record of observation with the CVs known today turns out to be always uncertain, a potential historical observation alone may not be relied on to draw conclusions on the evolution of binaries. The evolution scenarios should be derived from astrophysical observation and modelling only. 
\end{abstract}


%%%%%%%%%%%%%%%%%%%%%%%%%%%%%%%%%%%%%%%%%%%%%%%%%%

%%%%%%%%%%%%%%%%% BODY OF PAPER %%%%%%%%%%%%%%%%%%

	\section{Idea and Problem}
	During the last decades, there were some papers on the evolution of binaries based on the observation of a historical classical nova and its modern counterpart dwarf nova. \citet{pat2013}, for instance, titles `the oldest old nova and a bellwether for cataclysmic variable evolution' and uses the information, that `it has taken $\sim2000$ yr for the accretion rate to drop sufficiently' since the (observed) eruption to suggest some explanations for the `puzzle of CV evolution'. \citet{miszalski2016} writes similarly: `We match the DN to the historic nova of 483 CE in Orion and postulate that the nebula is the remnant of this eruption. This connection supports the millennia time-scale of the post-nova transition from high to low mass-transfer rates.' \citet{shara2017_nov1437} states in the abstract: `One scenario predicted that novalike binaries undergo a transition to become dwarf novae several centuries after classical nova eruptions' and that the few suggestions to prove this post-nova-to-dwarf nova metamorphosis with historical observation (\citep{shara2012}, \citep{shara2007}, \citep{pat2013}, \citep{miszalski2016}) lack robust and independent determinations of the date of the associated classical nova. This contribution in 2017 was ground breaking because of the astrophysical part: The results of determining the age of the shell and the proper motion of the associated CV fit sufficiently for conclusions on the evolution of binaries. However, we consider the identification with the Korean guest star in 1437 questionable and, therefore, not being able contribute reliably to the astrophysical problem. This leads to the question: How reliable are the other cases of identifications of historical novae with modern CV counterparts?  
 
 In this paper, we will discuss the most famous cases, i.\,e. the Nova 1437 in Sco (\citet{shara2017_nov1437}), the case of the strange star BK Lyn, and the discussions of historical observation of the eruptions which caused the shells of AT Cnc \citep{shara2012} and Z Cam \citep{shara2007}. The pattern of our discussion is always the same: We display the text of the original report according to modern translations, then we analyze and interpret the text in order to derive the position in the sky where the object appeared. After locating it, we compare our result with the position derived by other authors and evaluate the astrophysical consequences for the long-term behavior of cataclysmic variables (CVs) and the evolution of such binary systems.
 
	\subsection{Asterisms as positioning system}
	In the Far Eastern tradition, the designation of stars has always been a combination of the name of the constellation plus a number. For instance, the bright star Regulus ($\alpha$~Leo) has always been designated `Xuanyuan 1'. Only a few stars additionally have sound names that are more than a number. 

  In China the classical set of $\sim$1464 individual stars is grouped in $\sim$283 constellations (more detailed information on the number of asterisms per map, country and epoch compiled in \citet[e.\,g. p.\,93]{sunKistemaker}). Many of them can be traced back to the 1st and some even to the 4th century BCE. It is common sense that the Chinese Suzhou map (preserved on marble from 1247) displays the canonical system of constellations and their main stars. This marble map was copied to Korea in the 14th century. In Korea it is named \textit{Cheonsang yeolcha bunyajido}, the natural order of the cosmos and the regions they reign. The Korean marble plates, engraved in 1395 and stored near Pyongyang were destroyed during a later Chinese invasion but copies on other media survived. This map is a witness for the Chinese roots of the Korean celestial frame of reference. It is certain that the Japanese and Korean system of asterisms, lunar mansions and other divisions of the sky are copied from the Chinese one. The canonical system was developed during the Han dynasty (206 BCE to 220 CE) and especially as a consequence of a calendar reform in 104 BCE during the following decades in the 1st century BCE. The Han system contains asterisms of three `schools' \citep[chapter 2.3]{sunKistemaker}: the Shi Shi (92 constellations), the Gan Shi (118 constellations), and the Wuxian Shi (44 constellations). Shi Shi is the older system, developed between 481 BCE and 222 BCE but the original book was lost by Han time. Gan Shi and Wuxian Shi probably did not exist before the Han and are additions to the Shi Shi. These 254 constellations are the base for the canonical system developed thereafter. As far as we know, later additions did not change the earlier asterisms but only contributed more asterisms for the gaps in between them -- most likely to obtain better results for making calendars.  

	\subsection{Lacking Far Eastern star coordinates}
	However, there are two main difficulties in the interpretation of the celestial positions of stars: First, there are no complete lists of stellar coordinates before the 17th century CE (except the main stars) and, second, there are some uncertainties concerning the way of counting the stars within a constellation. First, the historical star catalogues provide only coordinates for the leading stars of the asterisms. There are plenty of lists of the scheme `asterism name, number of stars belonging to the asterisms' (e.\,g. also preserved as description engraved in marble next to the Suzhou map) accompanied by the coordinates\footnote{The coordinates are given as \textit{qu ji du} (distance to the pole, i.\,e. $90\degr\ -\delta$ declination), \textit{ru xiu du}. \textit{ru xiu du} is a sort of right ascension but measured per lunar mansion starting from the determinative star instead of a continuous measurement from the equinox. Additionally, there is \textit{huang dao nei wai du} (ecliptical latitude), cf. \citet[p.\,41]{sunKistemaker}.} of the main star. However, the other stars of the constellation are not listed with coordinates. Second, concerning the counting of the stars there are two conventions preserved (at least) since the Han: counting in the direction of right ascension (RA), $(a)$ starting from the border of the lunar mansion, or $(b)$ starting from the leading star. In most cases, the leading star is chosen on or near the western boundary of the asterism. Exceptions might be caused by the special astronomical or astrological relevance of another star. 

Consequently, as ($i$) there are exceptions from the rule of defining the main star, ($ii$) not all main stars are securely identified,\footnote{Many suggestions by different authors and their own -- most likely -- identification given in tables \citet[p. 47--52]{sunKistemaker}.} and ($iii$) there is more than one convention of counting the stars, a number of our modern identifications of the stars mentioned in the historical reports is uncertain (cf. e.\,g.  \citet{sunKistemaker}, \citet{pankenier2013}, \citet{bidaud}) or only likely (cf. our discussion of the nova 101). For our discussion of the few historical nova candidates announced above, we will now rely on the historical map (Suzhou) which is considered to preserve a common sense of the (almost?) canonical system. It is valid for the 15th and 17th century events and also gives very likely estimates for the events 101 CE and 77 BCE (as well as all other options discussed below). 

	\subsection{Resolving different dates} 
\begin{figure}
	 \caption{Difference in year numbering.} 
	 \label{fig:years} 
	 \includegraphics[width=\columnwidth]{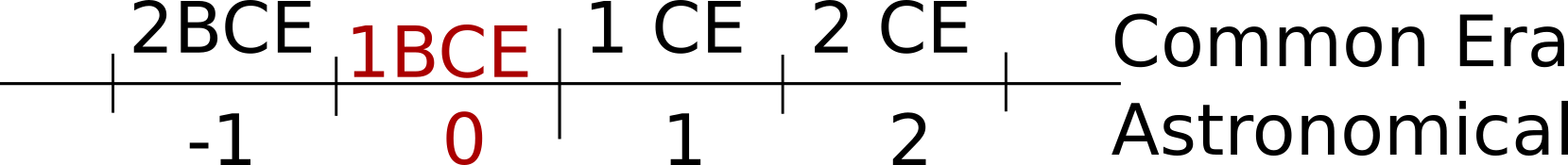} 
	\end{figure}

	Please remember, that the counting of years of the `Common Era' (CE) differs from the year number in astronomical year numbering because astronomical year numbering has a year number zero which is not the case for the common era (cf. Fig.~\ref{fig:years}). 
	
	\subsection{Technical explanations on plots of possible novae} 
	As displayed in the stemma in Fig.~\ref{fig:filters} at the end of this article, all cases of possible historical sightings of classical novae discussed in astrophysics rely on the list of possible novae by \citet{stephenson}: He himself copied this list in his later works (e.\,g. the often cited book on historical supernovae and novae, \citet{steph77}) and others also do, e.\,g. \citet{duerbeck}. In order to appreciate Stephenson's work, in our analysis we additionally used the lists of nova candidates which were published before this breakthrough article and, thus, were discussed in his book: Our plotted maps, additionally to Stephenson's suggested coordinates display the suggestions by \citet{hsi}, \citet{xi+po}, and \citet{pskovskii}. \citet{nickiforov} provides a new list of possible candidates and coordinates (maybe independent of Stephenson) which is unfortunately full of fundamental mistakes and misunderstanding which is why it is neglected by us. The maps of Fig.~\ref{fig:bkLyn}, Fig.~\ref{fig:zetScoSearch}, and Fig.~\ref{fig:nandou} are plotted in Bonne projection with Wolfram Mathematica 11.2 (2018/9).  

	\section{Case: BK Lyn as nova 101?} 
	A Chinese record in 101 CE is often regarded as classical nova eruption and the variable star BK Lyn is then considered to be its modern post-nova counterpart. This identification by \citet{hsi} and \citet{hertzog1986}\footnote{Hertzog (p.\,39) refers to the star as PG 0917$+$342. Those days, the star was not yet named because the discovery had been recent.} is the base for recent papers by \citet{pat2013} and \citet{kemp}. BK Lyn is a cataclysmic variable consisting of a white dwarf and an M-type star which permits dwarf novae (like U Gem) and sometimes even super-outbursts (subtype ER UMa). The subtype of ER UMa-like stars belong to the SU UMa class but reveal shorter cycle cadences of super-outbursts between 20 to 90 days. Within this period, they show rapid but small-amplitude normal outbursts ($\sim3$~mag) caused by low mass transfer rates within these systems.  

 However, this dwarf nova like behavior of BK Lyn was only seen during a short (unknown) time interval of a few years. More recent observations by AAVSO members show that the star has now returned to a more quiet state without dwarf nova outburst activity.

 Since the star has only $\sim15$ mag, it is not visible to the naked eye.

	\subsection{The historical report} 
   The historical record reads as displayed in Tab.~\ref{tab:event101}.
	\begin{table} 
	\caption{Three translations of the historical record for Nova 101. 'Xuanyuan' is the modern transliteration for the Chinese term formerly transliterated 'Hsien-Yuan' or 'Hsüan-yüan'. Thus, all three versions refer to the same asterism.}
	\label{tab:event101}
	\begin{tabular} {p{.27\columnwidth}|p{.27\columnwidth}|p{.27\columnwidth}| }\hline
	Xu et al. & Ho & Hsi \\ \hline
	Emperor He of Han [China], 13th year of the Yongyuan period, the 11th month, day yichou [2]. There was a small guest star in the space of the fourth star of Xuanyuan. Its color was bluish-yellow.
	& (\dots) a small guest star appeared at the fourth star of Hsien-Yuan. It was bluish-yellow in colour. 
	& 2d cyclical day, 11th month, winter, 13th year of Yung-yüan of later Han, small guest star around No. 4 of Hsüan-yüan, blue-yellow
	\\ \hline
	Hou Han shu, Tianwen zhi, ch. 21
	& HHS, WHTK, THHY,B,L,Hsi
	& Tung-han hui-yao, Hou-Han-shu, T'ung-k'ao
	\\ \hline
	\end{tabular} 
	\end{table} 
	
\begin{figure}
	 \caption{The Chinese asterism of Xuanyuan: left on traditional Chinese Suzhou map (black marble), right: the way of counting from west to east.} 
	 \label{fig:xuanyuan} 
	 \includegraphics[width=\columnwidth]{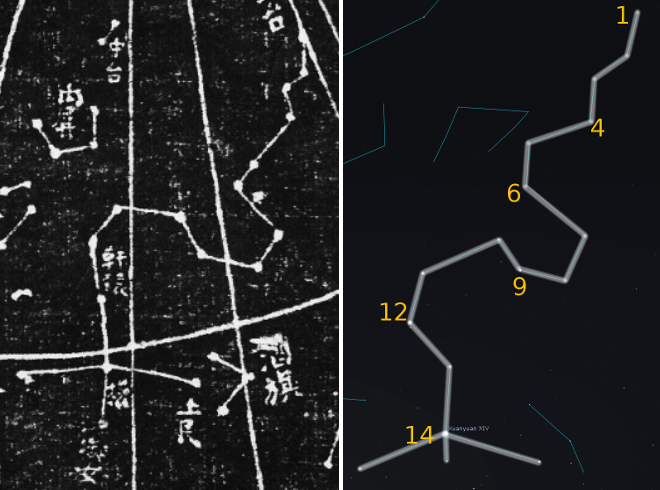} 
	\end{figure}

  The date given in the text translates to Dec 30th 101 CE. The given constellation Xuanyuan ($=$ Hsien-Yuan in old transliteration) is a long chain of stars from the vicinity of Regulus northwards to the area north of $\alpha$ Lyn. Fig.~\ref{fig:xuanyuan} shows a detail from the traditional Chinese Suzhou map and the common (modern) numbers for the stars in the chain.\footnote{The map is based on the information of the Hong Kong Space Museum where this information is no longer provided. However, it is distributed by Karrie Berglund of Digital Education Solutions, Inc. who also added it to Stellarium. The modern star numbers follow the historical tradition. In later Versions of Stellarium (e.\,g. in 0.19.$\ast$), more variants of the Chinese sky culture are added.}  
% \onecolumn
%\twocolumn 

	\subsection{Identification with BK Lyn} 
	There are two ways to interpret the standard counting of the stars in a constellation: $i)$ starting from the western end of the line and counting southwards or $ii)$ starting from the main star while sometimes it is unclear which one the main star is \citep[p.\,46--52, 67]{sunKistemaker}. In case of Xuanyuan, the most common counting starts from the upper end of the chain counting southwards. This counting is standardized today but also supported by the Suzhou map where the upper most star is at the RA-line delimiting the lunar mansion. Therefore, we arrive at $\alpha$~Lyn for the 'fourth star of Xuanyuan'.\footnote{This was suggested by Hsi (1957) under the name '40 Lyn' (equaling $\alpha$ Lyn), not discussed and not changed in Ho (1962), Hertzog (1986), p.\,39 endorses Hsi and misinterprets the '$\alpha$' in Ho's map on page 221 as '4'. However, since Ho very likely would have mentioned it, if he disagreed with Hsi like he did in other cases, we can assume that Ho also supports the identification of the fourth star with $\alpha$ Lyn.} Looking for cataclysmic variables in the vicinity of $\alpha$ Lyn, a check of the AAVSO VSX catalogue \citep{watson} in September 2018\footnote{\url{https://www.aavso.org/vsx/index.php?view=search.top}.} results in three CVs while only one of them is bright enough (brighter than 18~mag) to produce a classical nova visible to the naked eye. This only remaining candidate is BK Lyn (as suggested by \citet{hertzog1986}), a dwarf nova of ER UMa-type with a V-brightness varying between 14.3 and 16.5~mag and a period of 0.07498 days.  

	 \subsection{Consequences for the evolution of binaries} 
 As \citet{hertzog1986} refers to it as `one of the best-located nova candidates from Far Eastern records', this object is used for astrophysical conclusions. There are two publications with almost identical content: by \citet{kemp} in the proceedings of a backyard astronomers conference and \cite{pat2013} in MNRAS, while Kemp and Patterson have been co-authors for each other and most of the other co-authors are also the same. The abstract of the MNRAS-paper summarizes: 
 \begin{quote}
 ``Reviewing all the star's oddities, we speculate: (a) BK Lyn is the remnant of the probable nova on 101 December 30, and (b) it has been fading ever since, but it has taken $\sim$2000 yr for the accretion rate to drop sufficiently to permit dwarf-nova eruptions. If such behaviour is common, it can explain other puzzles of CV evolution. One: why the ER UMa class even exists (because all members can be remnants of recent novae). Two: why ER UMa stars and short-period nova-likes are rare (because their lifetimes, which are essentially cooling times, are short). Three: why short-period novae all decline to luminosity states far above their true quiescence (because they are just getting started in their post-nova cooling). Four: why the orbital periods, accretion rates and white dwarf temperatures of short-period CVs are somewhat too large to arise purely from the effects of gravitational radiation (because the unexpectedly long interval of enhanced post-nova brightness boosts the mean mass-transfer rate). And maybe even five: why very old, post-period-bounce CVs are hard to find (because the higher mass-loss rates have `burned them out').''
 \end{quote} 
 To evaluate these conclusions we have to understand human astronomical practice as well as the phenomenology of dwarf novae: \cite[p.~1910]{pat2013} point out that ``Somewhere between 2002 and 2005, the star faded sufficiently to allow dwarf-nova outbursts to occur.'' In this case, we seem to have observed the metamorphosis of a star. From the model-derived information that binaries with short orbital period $P_\text{orb}$ (like BK Lyn: $P_\text{orb}=0.07498$~days) show low mass transfers of $10^{-10}$~M$_\odot$\,yr$^{-1}$ they additionally conclude a time scale for the aftermath of the eruption of $1000-2000$ years and endorse the identification of BK Lyn with a (hypothetical) nova in 101 CE: ``In this scenario, BK Lyn is the product of a recent nova eruption -- probably, though not necessarily, the event of 30 December 101.'' This seems to fit the still high temperature of the white dwarf observed in UV and the star's light curve. However, for historical reasons it is impossible to decide from observation whether the recent metamorphosis of this system was unique. 

 First, even at Far Eastern courts we cannot be sure that the astronomers surveyed the sky \textit{systematically}. The preserved records are preserved for their divinatory importance only and they are abbreviated notes in chronicles. In many cases, the original diaries of astronomers are lost or did not exist; surviving records are preserved in later technical treatises, biographies, and annals included in large-scale historiographical projects. The astronomers certainly pursued continuous observations and wrote down every anomaly they happen to become aware of. But they also certainly did not survey certain stars. As their goal were divinatory predictions for the sake of the empire, they derived an omen from the appearance of a new object and then their job was done. Guest stars were especially important for this purpose because they were not regular and not predictable. Although during the Eastern Han period, the Chinese made advances in predicting eclipses because of their better understanding of the physical process causing it, they did not think about stellar physics as cause of guest star appearances and, therefore, did not monitor each object after its appearance or even after its disappearance. 
 
 Second, typical amplitudes of dwarf novae are of the order of 6~mag. Thus, a dwarf nova of BK Lyn ($\sim15$~mag) would brighten the object to a magnitude of 10 to 8~mag which is still beyond the eye's detection limit. No astronomer on Earth would have been able to witness a dwarf nova outburst of BK Lyn even if they had monitored this object (or area).

	For the abstract cited above, this has the following consequences: Assuming the hypotheses being correct that the guest star 101 is related to BK Lyn and, thus, this guest star designates a nova, this does not lead to sentence (b) as conclusion but sentence (b) comes out of the model \textit{only}. If the star produced a classical nova in $+101$ and has not been observed since then, this does not mean that it has taken roughly 2000 years to permit dwarf-nova eruptions. It only means that there are no records of the star being observed (randomly) by humans (who did not monitor it). Hence, the single observation of a guest star does not contribute to the model of evolutions within binary systems but this model of mass transfers in cataclysmic binaries is needed to suggest the star BK Lyn as possible cause for the guest star. Concerning the evolution of the binary system, the central question ``What can we derive from historical guest stars?'', the unsatisfactory answer is: Little concrete information can be derived from historical observations of guest stars. 
	
%	\onecolumn
 	\begin{figure*}  
 	\caption{The filled circles indicate the coordinates of earlier interpreters of the text. They all published lists of point coordinates and we apply a margin of error of $5\degr$ (because in their lists all declination numbers end with 0 or 5) obtaining these filled circles (all equinox 1950). The bold dark gray circles (rings) indicate our own interpretation of the position of the appearance: The text can be read as suggesting a guest star next to $\alpha$ Lyn or next to $\zeta$ Leo. The red star and green diamond symbols mark CVs in our circles according to the AAVSO VSX database (2018). Star 2 marks BK Lyn; for further information see Tab.~\ref{tab:101CVs}.} 
	 \label{fig:bkLyn} 
	 \includegraphics[width=\textwidth]{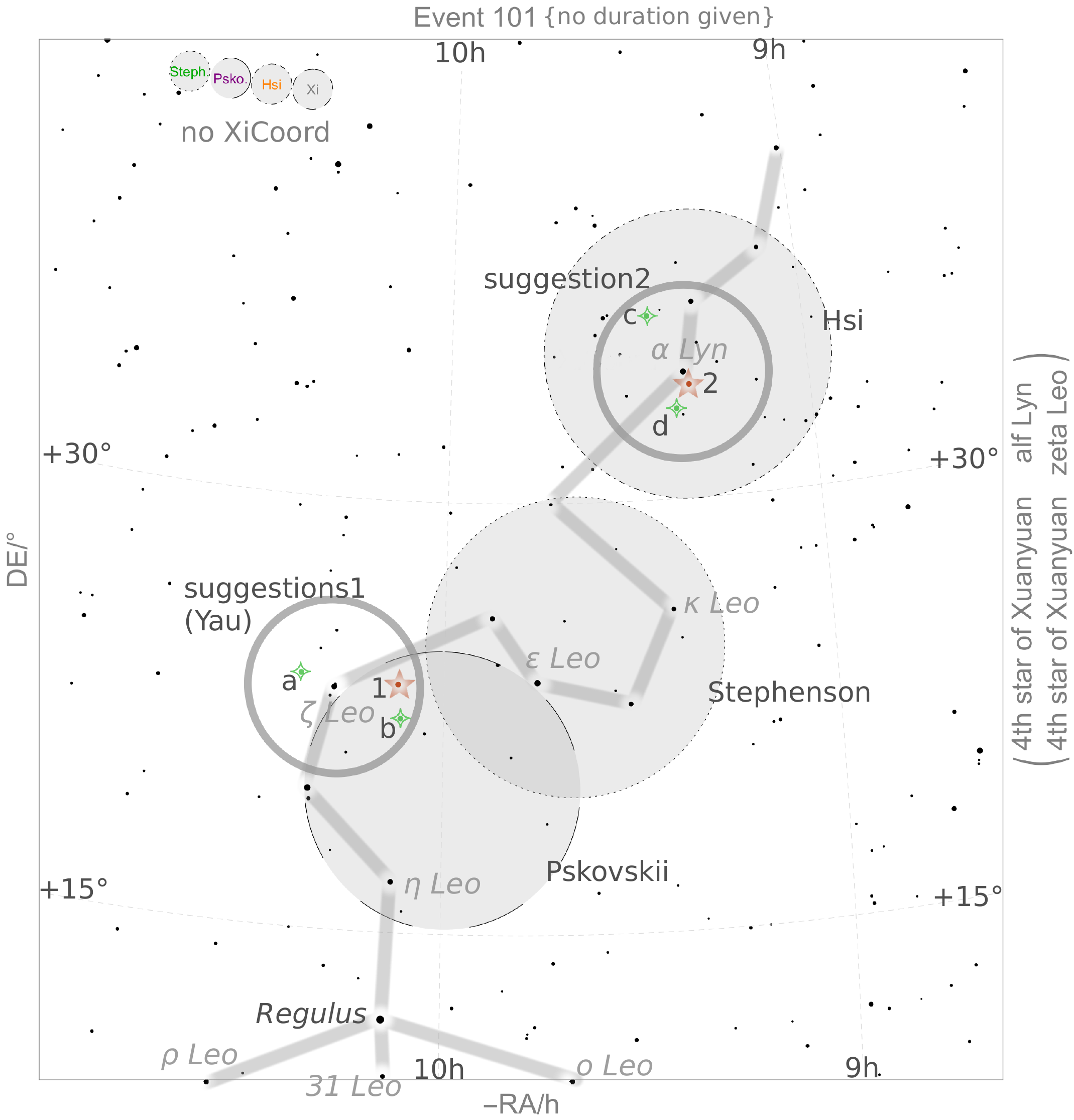}
	\end{figure*} 
	\begin{table*}
	\caption{Cataclysmic (and no symbiotic) binaries in our fields for the VSX query. $\star$-symbol for CVs which are normally at least 18~mag or brighter, $\diamondsuit$-symbol for fainter CVs.}
	\label{tab:101CVs}
	\begin{tabular}{ll|ll} 
	\multicolumn{2}{c}{Legend for VSX output:}\\
	$\star$ 1 & SDSS J100658.40+233724.4 & $\star$ 2 & BK Lyn \\\\
	$\diamondsuit$ a & IK Leo& $\diamondsuit$ c & YZ LMi \\ %\hline
	$\diamondsuit$ b & CSS 161011:100629+222644 & $\diamondsuit$ d & SDSS J092229.26+330743.6 \\
	\end{tabular} 
	\end{table*}
%\twocolumn
	\subsection{Uncertainty of the interpretation}
	Finally, we have to consider the reliability of the report. The excerpts cited above are grabbed from several chronicles, compiled in different epochs. The name of the Emperor is given as `He of Han', which is the posthumous name of Liu Zhao. The report is, therefore, not a live-reportage from astronomers to their emperor but it is a sentence from the corpus of astronomical records which had been interesting enough for a chronicler to mention it in the story he wants to tell. In this context, we have to recall that historical chroniclers cannot follow modern rules of historical research but are aiming to tell a story for political reason. That means, they select astronomical (and other) events and put it in an appropriate frame which emphasizes the goal of communication (e.\,g. the correctness of a political decision or the explanation for a lost battle). Considering the role of astronomy in the Chinese culture, we can be sure that the astronomer's original report was telling the truth and that the chronicler did not change anything. However, the astrological relevance of an event derives from the appearance of the guest star alone and not from the physics causing it. Possible changes in colour, size, or brightness had either not been observed or the reports are simply not preserved. 
	
	As no tail or movement of the star is reported, we do not have any hint to interpret it as comet or meteor. Thus, the object might have been a nova. However, it is neither given how long the guest star lasted nor any further information. That is why we cannot be sure that there had not been a tail: Maybe, there was a tail and a movement which is lost in the abbreviated excerpt of the preserved fragment of text after passing the filters of divination and chronicle relevance. Hence, we are not even sure that the event was a nova -- although the description and the vicinity of a strange cataclysmic binary suggest this interpretation. 

	Concerning the position of the event (and the vicinity of a strange CV), \citet{stephenson}\footnote{In \citet{steph77} the coordinates are not changed.} and \citet{pskovskii} suggest different positions (see Fig.~\ref{fig:bkLyn}). Pskovskii possibly starts the counting of stars from the main star of the asterism, Regulus\footnote{In this cases, everybody agrees to identify Regulus ($\alpha$ Leo) as main star of Xuanyuan: cf. \citet[p.49]{sunKistemaker}.}, which leads to the interpretation of $\zeta$ Leo to be the fourth star. While this agrees with Stephenson's doctoral fellow Yau\footnote{In his cumulative dissertation on PDF-page\,236, which has the (unsystematic) page number 115.} who relies on \citet{liu} for the suggestion of counting from the main star (Regulus), Stephenson's own interpretation of the position differs from both versions, i.\,e. from the Pskovskii/Yau suggestion next to $\zeta$~Leo as well as from the Hsi/Hertzog one next to $\alpha$ Lyn (cf. Fig.~\ref{fig:bkLyn}). \citet{stephenson} suggests the position of the nova between $\epsilon$ and $\kappa$~Leo and does not give a reason. Probably, he overlooked the term `fourth star' and simply took the middle of the constellation of Xuanyuan. This is likely, because he gives the same coordinates for the event in $70$~CE where the historical report says only `in Xuanyuan'. 

\begin{figure}
	 \caption{Screenshot from the table in Liu's paper with our highlighting of terms (blue: constellation name, yellow star name): The first column is just the line number in the table, the second column gives the modern IAU star name. The third column identifies this with the modern Chinese star name (such as "`Xuanyuan 14"'). The fourth column adds two historical names for $\alpha$ Leo, `bright star of Xuanyuan' and `Xuanyuan, star of the queen', and one for $\rho$ Leo, `Xuanyuan, left star of the people'.} 
	 \label{fig:xuanyuanStars} 
	 \includegraphics[width=\columnwidth]{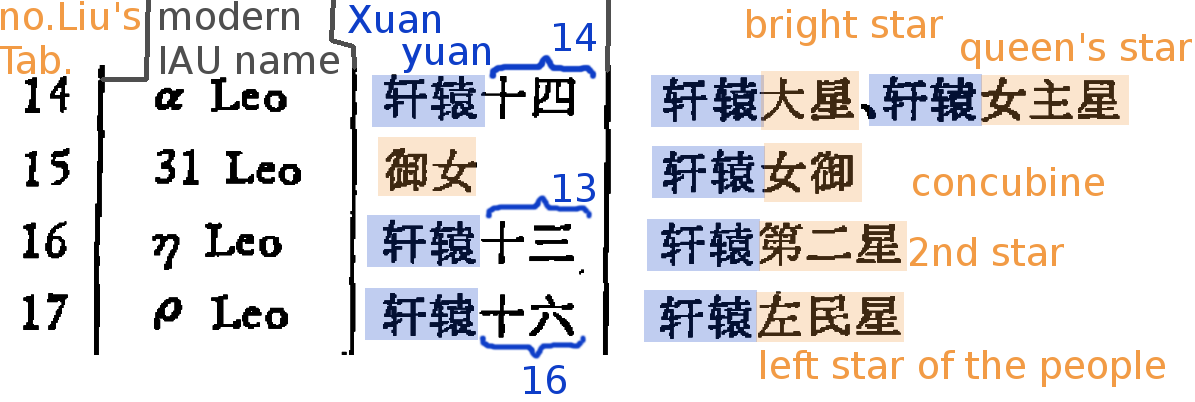} 
	\end{figure}

 The base for the identification of $\zeta$~Leo as the fourth star is the work by \citet{liu} concerning the names of 50 zodiacal stars used in the 4th to 6th century (roughly half a millennium after this report but star names seem to be highly conservative from around 100 BCE). Studying the positions of planets, Liu derives that the `second star of Xuanyuan' was $\eta$ (cf. table in Fig.~\ref{fig:xuanyuanStars}) while for $\alpha$, $\rho$ and 31~Leo proper names were in use. However, Liu's method is only useful for stars with very small distances from the ecliptic where the planets pass by. As there are sometimes multiple names for a star, it is possible that their counting depends on the astronomical purpose. Observing the planets forces to consider only the stars $\alpha$ Leo, 31 Leo, and $\rho$ Leo (named `star of the queen',`the concubine', and `the star of the people', respectively) because only these can be touched or covered. $\eta$ Leo has an ecliptical latitude of $4\degr\,45^\prime$ and, thus, can be covered or touched only by the Moon (which is mentioned in Liu's paper). All the stars north of $\eta$ Leo in Xuanyuan are not relevant for zodiacal astronomy. That is why, this astronomical purpose suggests to start the counting from Regulus (main star) or $o$~Leo (the western most of the zodiacal part of Xuanyuan; not in Liu's table). For non-zodiacal purposes we still have to consider the whole chain of Xuanyuan.

	A division of a huge constellation in subgroups with separate numbering is also well proved in Liu's table: The asterism of The Well now containing eight stars provides seven star names for this analysis grouped in four subgroups. 
	
	In contrast, the numbering of the stars in the asterism of the Net (also in Liu's table) in Taurus is the same as today, starting with the northern most star and not with the western most (culminating early by some 25 minutes). This supports our interpretation that multiple ways of counting the stars had been in use and without knowing the observer and his cultural and scholar background as well as the purpose it is difficult to find out the exact meaning of such terms. Thus, in our Fig.~\ref{fig:bkLyn} we defined two circles and suggest them to search nova candidates within: ``suggestion\,1'' is the circle around $\zeta$~Leo, the fourth star in case of counting from Regulus northwards (which certainly had been the case for purposes of zodiacal astrology) while ``suggestion\,2'' is the circle around $\alpha$~Lyn in case of counting the stars from the westernmost right ascension of the chain of stars southwards (which is the usual practise of counting stars). 
	 
	 In case of the usual counting and identifying $\alpha$~Lyn as `fourth star of Xuanyuan', we obtain BK Lyn as best CV candidate to have caused a classical nova. In case of the zodiacal counting and identification of $\zeta$~Leo as `fourth star of Xuanyuan' the only CV which is bright enough to be a good candidate to have caused the event would be the star SDSS $J100658.40+233724.4$, an eclipsing UG binary with a period of 0.185913 days. Its brightness is normally between 18 and 17~mag, while outbursts can brighten it to less than 15~mag and eclipses take it to $\sim20$~mag. Although the latter identification might appear less likely (for astrophysical as well as for historical reasons), its existence weakens the contribution of historical guests stars to modern research.
	 
	 \subsection{Conclusion}
	 It is uncertain but the likeliest possibility that BK Lyn caused the guest star reported in 101 CE if this guest star was a classical nova. As this is not certain, the transient of 101 CE could also be something else but a classical nova. 
	 
	\section{Case: Nova 1437 Sco} 
 In 2017 Shara et al. presented the discovery of a nova shell in Scorpius. Within the shell, they found the cataclysmic variable 2MASS J17012815-4306123 with strong emission lines and the X-ray source IGR J17014-4306 (probably the hot spot of its disk). Tracing back the centre of the shell and the CV to the year 1437 by measured proper motion leads to a match and, therefore, the conclusion that the CV 2MASS J17012815-4306123 caused the shell. 

   The parallax measured by Gaia is $0.9602\pm0.0483$~mas leading to a distance of the CV of $1.041\pm0.0523$~kpc. That means, the nova shell with a radius of about 50~arcsec or 0.82 light years extended from its centre to the current state within 579 years from 1437 to 2016 which allows interesting insights in this evolution.  

	First doubts occur since the proper motion of the CV measured by Gaia DR2 \citep{brown} are only a third of the proper motion measured by Pagnotta in \citet{shara2017_nov1437} and used for the computation of the position of the star in 1437:

   \begin{tabular} {lll} 
				& $\mu_\alpha $/ mas/yr		& $\mu_\delta$/ mas/yr   \\ 
   Shara et al. '17	& $-12.74\pm 1.79$ 			&  $-27.72 \pm 1.21$   \\ 
   GAIA 2018		& $-4.23\pm0.093$ 			& $-9.644\pm 0.073$  \\ 
   \end{tabular}

  At first glance, the new measurements seem to suggest that the CV proposed in 2017 does not fit the shell. Although a factor of 3 has only an effect of a few arcseconds for the matching of star position with shell center (likely causing only changes of error bars), this contradiction surprises because the parallax derived in \citet{shara2017_nov1437} uses a baseline of 93 years and, thus, seems to be reliable. In contrast, for reasons of the variability of a CV and the nebulosity around this particular star, it is well possible that the Gaia data in this case is erroneous; \citet{shara2017_nov1437} might be right with the determined proper motion. However, this case of ancient nova-modern CV identification provokes revisions -- at least from the historical point of view.  
  
	\subsection{The historical report} 
 The historical record is from Korea and we cite it from \citet[p.\,143]{xu2000} and \citet[No.\,(508), p.\,202]{ho} in Tab.~\ref{tab:records1437}. 
 	\begin{table}  
	\caption{Translations of the historical record for Nova 1437.}
	\label{tab:records1437}
	\begin{tabular} {p{.36\columnwidth}|p{.36\columnwidth}|p{1cm}|}\hline
	Xu et al. & Ho & Hsi \\ \hline
	19th year of King Sejong, 2nd month, day yichou [2]. A guest star first appeared between the second and third stars of Wei [LM 6]. It was nearer to the third star and separated from it by about half a chi. It lasted 14 days in all.
	 & On an i-chhou day in the second month of the 19th year a `guest star' appeared between the second and the third star of the Wei (sixth lunar mansion) and went out of sight after 14 days.
	 & --
	\\ \hline
	Yijo sillok, Sejong sillok; ch. 76
	 & Ch\u{u}ngbo Munh\u{o}n Pigo ch. 6.
	 & --
	\\ \hline
	\end{tabular} 
	\end{table} 

 The date translates to March 11th 1437. The asterism `Wei [LM6]'\footnote{The text only gives `Wei' and usually in case of no further specification the asterism is meant. This asterism also names a lunar mansion. Since there are many `wei' in Chinese and actually three constellation names are transliterated like this, some scholars (including Xu et al.) distinguish between them by adding the number of the lunar mansion (without implying the LM is meant).} is one of the oldest ones and defines a lunar mansion(LM). Its depiction on the Chinese Suzhou map is shown in Fig.~\ref{fig:wei6maps}. 

	\subsection{Identification of the shell with nova 1437} 
 In their paper, \citet{shara2017_nov1437} refer to two (not cited) lists from the Chinese Yuan dynasty with different order of stars: One of them counts $\mu, \epsilon, \zeta\dots$ while the other one counts $\epsilon, \mu, \zeta\dots$, starting from the main star and the first in the chain, respectively. However, both lists consider $\zeta$ as third star and since $\mu$ and $\epsilon$ both are laying north of it, according to the text, the guest star began to be seen north of $\zeta$~Sco -- but the newly discovered and age-dated shell is not. It is located roughly in the middle between $\eta$ and $\zeta$ Sco. 
 
\begin{figure} 
  \caption{left: Korean, right: Chinese historical map with modern star names added. Note the red line (RA of the lunar mansion) in the Korean map crossing $\mu$ Sco indicating the first star of the asterism.}
  \label{fig:wei6maps}
  \includegraphics[width=\columnwidth]{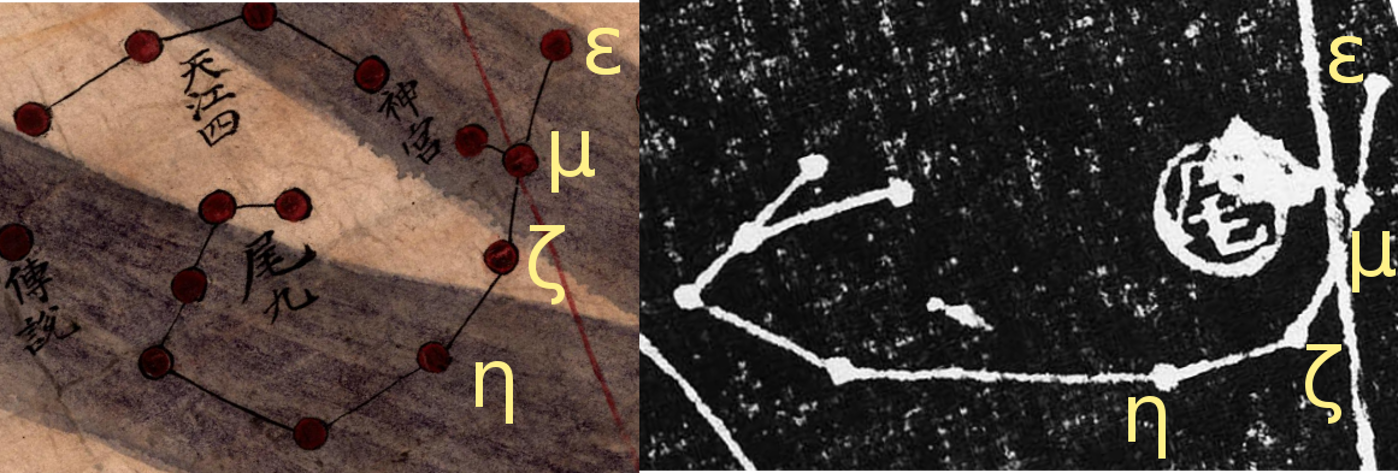}
 \end{figure}

 Stephenson, in all his publications of nova catalogues since 1976 locating the event between $\zeta$ and $\mu$~Sco, now changes his interpretation of star counting and suggests to count from the main star in a circle in order to obtain $\eta$ as third star -- but without any historical base, against everything we know about the purpose of star counting.\footnote{The counting usually follows increasing right ascensions, and, therefore, the order of rising (hour angle) of asterisms which is appropriate for historical measurements of time (cf. previous chapters).} Interestingly, \citet{shara2017_nov1437} cite two Chinese lists of stars from Yuan dynasty (ending $\sim$100 years before the considered guest star) and does not take into account the preserved \textit{contemporary} Korean map (cf.~\ref{fig:wei6maps}) which had certainly been the base for Korean astronomers. The Korean map is preserved in the state of 1395 and already registers the 'added star' next to the determinative star of the lunar mansion. The Koreans likely did not change the Chinese counting along the line from north to south and west to east since their chain of stars shows the same pattern. Consequently, there is no need for us to suggest another counting than given in the historical lists. 
 
 However, one could impute the assumption to Stephenson that the naked eye observers or (more likely) the chroniclers while including the fragment in their texts mixed up directions to stars. There are numerous historical cases from several cultures showing that descriptions of positions are sometimes turned around and the preserved text gives `east' instead of `west' or `north' instead of `south' (cf. Hipparchus's alignments to prove the stability of patterns of fixed stars \citep[chap. 2.5.2]{smh2017} as well as Chinese descriptions of the supernova 1054 with regard to the star $\zeta$~Tau). Yet, any mixing up of the stars north and east of $\zeta$~Sco is especially unlikely because the straight line of $\epsilon, \mu, \zeta$ Sco is almost orthogonal to the line $\eta,\zeta$~Sco and the first one marks a line of right ascension. Thus, although it might be easy to mix up or intensionally exchange $\epsilon$, $\mu$, and $\zeta$ Sco for counting the stars, an exchange with $\eta$ Sco appears unlikely. We recommend the original interpretation of the position of the observation accepting that it does not fit the CV in the shell suggested by \citet{shara2017_nov1437}.
 
	\subsection{Consequences for the evolution of binaries} 	
 Modern observations (20th and 21st century) characterize 2MASS J17012815-4306123 as a CV with (now) low accretion rate that permits dwarf novae and the shell witnesses a classical nova eruption some centuries ago. That this CV fits the observed shell appears almost certain. For a classical nova, the accretion rate should have been high those days. Therefore, the system has been transformed from high to low accretion rates during the centuries in between. Since such an evolution is a prediction of the hibernation scenario, \citet{shara2017_nov1437} conclude in favor for this model. From the physical point of view, this logic appears coherent -- but what does it mean for our central question of the relevance of historical guest stars?
 
In their paper, \citet{shara2017_nov1437} use the identification of a CV in Scorpius with the guest star Nova 1437 identifying this observation as the classical nova. Because they found dwarf nova eruptions only on Harvard plates from 1938 to 1942 (with Harvard plate archive (DASCH) going back to the 19th century), they resume the interval of 500 or 600 years (generally spoken) to transform into a dwarf nova. This conclusion is critical because of the uncertainty of both fixed points: First, due to the uncertainty of the historical counterpart the date of the classical nova is not sure. Second, even modern observatories have gaps in monitoring projects, changes of environmental conditions,\footnote{e.\,g. Harvard Boyden observatory, located in Peru from 1889 to 1927 and in South Africa afterwards (it was moved).} changes of research focus, and normally they did not continuously survey the whole sky. Thus, the finding of dwarf nova eruptions on plates dating around 1940 does not mean that there were no dwarf novae eruptions before that date. It only means that no observations of the object or its outbursts are reported from earlier time.
%
%Assuming the correctness of the identification of the nova shell with the event in the historical Korean record, the conclusion on the time scale of the post-nova-to-dwarf nova metamorphosis is not allowed: They resume the interval of 500 or 600 years to transform into a dwarf nova because they found dwarf nova eruptions only on Harvard plates from 1938 to 1942. 
% To evaluate this hypothesis, we have to recall that the region of the Scorpion's tail is at southern declinations below $-34\degr$ and, therefore, cannot be observed at Harvard. The Harvard plate archive (DASCH) also contains data from 27 other observatories, some in the southern hemisphere (e.\,g. Peru) but their data starts at different historical times (earliest around 1900, but some only in the 1930s). Plate A12425 originates from Harvard Boyden observatory, located in Peru from 1889 to 1927 and in South Africa afterwards (it was moved). Like in this case, also the other observatories have gaps in monitoring projects, changes of environmental conditions, changes of research focus, and normally they did not continuously survey the whole sky. Thus, the finding of dwarf nova eruptions on plates dating around 1940 does not mean that there were no dwarf novae eruptions before that date. It only means that no observations of the object or its outbursts are reported from earlier time. 
 
 Assume the transformation time from classical nova to dwarf nova was only 40 to 60 years like in case of Nova 1960 Her (=V446 Her), cf. \citet{honeycutt}: Since the object's $g^\prime$ magnitude of 16.4~mag (cf. VSX catalogue, last checked November 2018) is far below the eye's detection limit (even in a dwarf nova peak of 2 to 6~mag) nobody would have recognized it even with systematic monitoring which certainly did not exist until the last few decades. 13 further cases of novae from 1862 to 1969 are presented in \citet{vogt}. They all show dwarf nova outbursts although their classical novae happened only $\sim150$ to $\sim50$ and not thousands of year ago. In their paper, \citet{shara2017_nov1437} note that the star brightened by 2 -- 4 mag during the 1930s; the VSX database reports magnitudes between 18 and 12 mag for this eclipsing binary. Hence, nobody would have recognized a dwarf nova outburst of this object e.\,g. in the 1480s or even in the 1580s. An evolution scenario like this, however, still allows conclusions by \citet{shara2017_nov1437} in favor of hibernation. Yet, the contribution of a historical guest star for the argument appears questionable. 
	
	Although the coincidence of the positions of the star and the shell found by \citet{shara2017_nov1437} seems to suit the guest star in 1437, we prove the given identification to be doubtful and the (recent as well as older) historical observations to be certainly not continuously monitoring. Hence, we recommend caution in drawing any conclusion for the evolution of close binary systems. 
 
	 \subsection{Critical discussion of alternative interpretations}
\begin{figure*}
  \caption{Until 2017 all authors of ancient-nova catalogues (i.\,e. \citet{stephenson}, \citet[p.\,46--49]{steph77}, \citet{pskovskii}, \citet{xi+po}, \citet{duerbeck}, where \citet{steph77} and \citet{duerbeck} rely on or even copied \citet{stephenson}) suggest a position between $\zeta$ and $\mu$ Sco. Our independent analysis leads to the same suggestion. Looking for CVs in a 3\degr\ radius around $\zeta$ Sco (as done for the paper \citep{shara2017_nov1437}) we found many objects and those of them plotted with $\star$ symbol into this map are bright enough to become naked-eye visible during a classical nova. Yet, we admit the guest star might have appeared more north and within the circle of Stephenson's original suggestion in a field we marked with a rectangle. We'll analyse this in a future paper on `A new approach to generate a catalogue of potential historical novae' \citet{hoffmannVogt2019}.}
  \label{fig:zetScoSearch}
  \includegraphics[width=\textwidth]{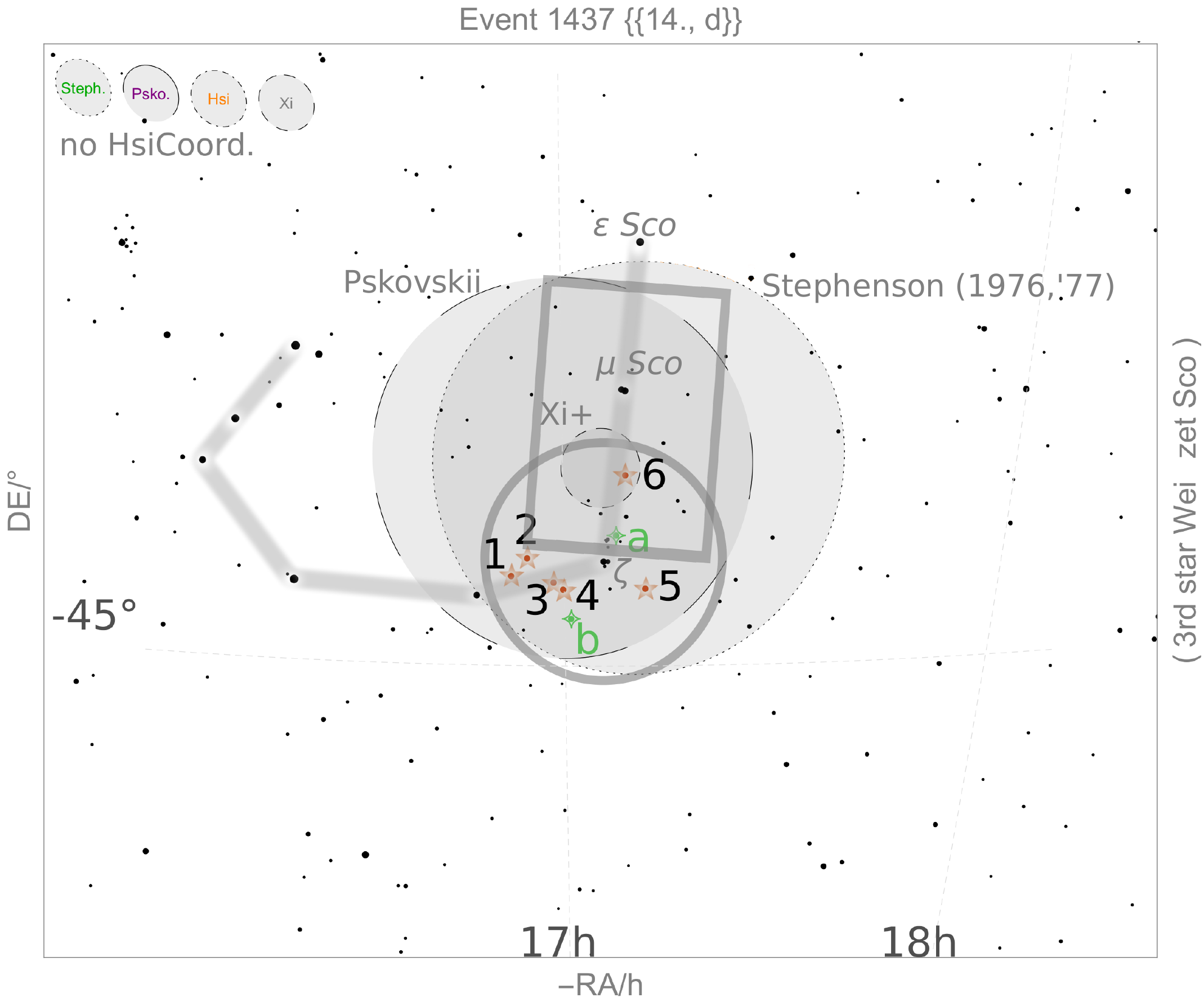} 
 \end{figure*}

 As we do not see a reason to change the order of star counting within the asterism, we keep identifying $\zeta$ Sco as third star of Wei and are looking for post-nova cataclysmic variables in the vicinity. In contrast to the claim of \citet{shara2017_nov1437} that there are no CVs around the given position close to $\zeta$ Sco, our plot (cf. Fig.~\ref{fig:zetScoSearch}) displays the resulting eight CV positions as output of the AAVSO VSX database (last query November 2018). Six of these eight CVs are usually brighter than 18~mag and, therefore, marked by us with a $\star$-symbol in the map. From these six candidates especially the cataclysmic variable VVV-NOV-017 is exactly at the position described in the historical report. 

 \textbf{From the astrophysical point of view:} However, there are several reasons against this object being a proper candidate: 
\begin{itemize} 
 \item This object was classified as nova candidate only in 2012 and because of a classical nova eruption \citep{saito2016}. The classification as classical nova might be questionable (observed was only 3.6~mag in brightening in the infrared) but let us consider all possibilities. If the recently discovered IR brightening really was a nova outburst, and if this system also caused the event visible to naked eyes (not IR) in 1437, it would have had two nova outbursts within 600 years which could possibly classify it as a recurrent nova.
 \item \textbf{magnitude arguments}:
  \begin{itemize} 
	\item The normal brightness of VVV-NOV-017 is roughly 18~mag. With regard to the possible and typical amplitudes of novae, this is at the faint limit for a candidate to be able to brighten to a magnitude visible to the human eye. Such an object should have outburst amplitudes of 12 or more magnitudes to be visible to the naked eye (which is possible but not usual for typical classical novae). In this case, the object should preferably provide higher amplitudes to become visible in front of the bright clouds of the Milky Way. This makes it a highly unlikely candidate. 
	\item If it was a recurrent nova, this is even more unlikely, because typical outbursts of recurrent novae known from the 20th century show smaller amplitudes than classical novae.\footnote{Vogt's contribution in \citet[p.\,239]{bode} states amplitudes between 8.1 and 10.1~mag for recurrent novae.} Such a system cannot reach brightnesses for naked eye observation. 
	\item The observed range of brightnesses of this system during the last years is between 18 and 14.3 mag. If it was a recurrent nova, the peak magnitudes would likely be similar. Since the nova eruption in 2012 was not visible to the naked eye, the eruption some centuries ago also would not have been. 
 \end{itemize} 
 \item \textbf{peak duration argument:} Additionally, typical peak durations of the known recurrent novae last longer than 14 days (but several months). On the other hand, all classical novae could be recurrent on large timescales (considering centuries or millennia) which is why this argument does not completely exclude the candidate but makes it unlikely. 
\end{itemize}
 The database entry of nova VVV-NOV-17 between the second and the third star of constellation Wei in Sco can, therefore, not be supported as a candidate to have caused the appearance in 1437. As we additionally expect (from observation as well as from theory) historical outbursts of (recurrent) novae should reach similar peaks as the presently observed ones. Thus, the physics of the event registered in 2012 does not suggest it as a candidate for the cause of the appearance -- although it fits the reported position. As suggested in the discovery letter \citep{saito2016} follow-up observations of this target are recommended to make sure that VVV-NOV-17 was a nova at all.  
  
   \textbf{From the historical point of view:} Yet, it remains unclear whether or not the appearance in 1437 really has been a classical nova eruption (or any outburst of a symbiotic star). The historical report gives only the fact that something appeared and does not describe the appearance by colour, brightness, or changes. It only mentions its duration of 14 days. Interestingly, 14 days later, on March 25th, two days after full moon, the moon (illumination still $\sim90$ per cent) approached the lunar mansion of Wei (crossing it on 26th) and was in close conjunction with Jupiter (occultation unobservable during the daylight hours). From the historical (divinatory) point of view, we could speculate whether or not the given duration in the chronicle is really based on observation of the transient or a result of interpretation of the observational notes by a non-astronomer. In the diary (if it existed) it could have been two items `[date 1] appearance of new star in Wei' and `[date 2] Jupiter Moon conjunction next to Wei' without any relation. It occurs well possible that the conjuction (or occultation) was considered to replace or reduce the guest star's influence with regard to divination. This could have led to the preserved text where a duration [of the appearance in the sky or of its divinatory regime?] is given but not a date of disappearance. This duration could possibly refer to the divinatory influence of the guest star and not necessarily to its visibility. We should not overestimate the observational information in the record.
%\onecolumn
 \begin{table*}
	\caption{These objects are the result of a query for cataclysmic variables and symbiotic stars in the AAVSO VSX database. The red star symbols indicate objects that are usually at least 18~mag or brighter while the green diamonds mark the fainter objects that are (for their faintness) unlikely able to outburst up to visual magnitudes visible to the naked human eye. Although it is written in the text that the appearance was seen north of $\zeta$~Sco, we made this query exactly around this star because this way we would `find' Shara's CV, too (as test of the method). Likely candidates (of our opinion) are only objects in the northern half of our circle such as $\star$~6 and $\diamondsuit$ b or even further north within the area marked by the box (for more information see our further paper \citet{hoffmannVogt2019}.}
	\label{tab:1437CVs} 
  \begin{tabular}{ll} 
	\multicolumn{2}{c}{Legend for VSX output:}\\
	$\star$ 1 & V0992 Sco \\
	$\star$ 2 & VVV-NOV-009\\
	$\star$ 3 & GDS\_J1701281-430612 $=$ 2MASS J17012815-4306123 in \citet{shara2017_nov1437}\\
	$\star$ 4 & VVV-NOV-008\\
	$\star$ 5 & N Sco 2018 No. 2 \\
	$\star$ 6 & VVV-NOV-017\\
	$\diamondsuit$ a & Gaia16avw\\
	$\diamondsuit$ b & Gaia18cbo\\
  \end{tabular}
  \end{table*}
%  \twocolumn

 Even if we take the given duration (of the appearance) for granted: Since the historical record only reports an appearance and does not report anything else but a duration of 14 days it is not even sure that this object was caused by a star. The term `guest star' might suggest a point-like source and no movement or tail is reported but that does not necessarily mean that there was no movement or tail. It could also have been a small comet or -- if it was caused by a fixed star -- any type of outburst stars can perform (e.\,g. merger, eruptions, explosions, flares\dots\ hypothetically even microlensing). 

  \subsection{Conclusion}
 The identification of the Korean guest star in 1437 with the the CV and its shell found by \citep{shara2017_nov1437} appears doubtful: The shell does not fit the position and at the position there is no appropriate CV. 

 For our central question, which information we can derive from reports of guest stars, we remain desillusioned: From the information of mere existence of a guest star we can only derive an appearance and neither a light curve nor any other type of phenomenological description. The identification of the historical counterpart of a modern shell and CV is, again, so weak that further research in astrophysics could help to clarify the identification but a usage of the historical report for the astrophysics is doubtful. 
		
	\section{Case: AT Cnc} 
 AT Cnc is a Z Cam-type star with a V-brightness between 15.8 and 12.5 mag and a period of 0.2016 days, discovered in 1968. In 2012 Shara et al. published a 3\arcmin-shell discovery around the object \citep{shara2012}. In 2017, Shara et al. published their analysis of the kinematics and the resulting age of the shell. There is one version of this publication with only the kinematics and without mentioning any historical case \citep{shara2017_ATcnc} and one version with a discussion of the historical appearance with co-author Stephenson \citep{shara2017_ATcnc_steph}. The result of this discussion is no certain suggestion of a historical counterpart. The result of the astrophysical measurements and analysis is an age of $330^{+135}_{-90}$~yr. Hence, the classical nova should have happened between 1552 and 1777 CE. 

	\subsection{Is there an identification of AT Cnc with a historical event?} 
 As AT Cnc is a star in Cancer \citep{shara2017_ATcnc_steph} scanned the historical records for a guest star in Cancer in this period and end their abstract with the statement `We conclude by noting the similarity in deduced outburst date (within a century of 1686 CE) of AT Cnc with a `guest star' reported in the constellation Cancer by Korean observers in 1645 CE.' The text of the Korean record is reproduced in Tab.~\ref{tab:eventATcnc}.
 	\begin{table}  
	\caption{Translation of the historical record for the appearance in 1645. Ho and Hsi do not register this because their collections end in 1600 (shortly before the invention of the telescope).}
	\label{tab:eventATcnc}
	\begin{tabular} {p{.6\columnwidth}|p{1cm}|p{1cm}|}\hline
	Xu et al. & Ho & Hsi \\ \hline
	King Injo of the Yi Dynasty [Korea], 23th year, 2nd month. A large star entered Yugui.
	 & -- 
	 & --
	\\ \hline
	Chungbo munhon pigo, ch. 6
	 & --
	 & --
	\\ \hline
	\end{tabular} 
	\end{table} 
    
\begin{figure}
   \caption{Historical maps: left: Korean (original on parchment/ leather), right: Chinese (original on black marble). The asterism `Yugui' is the little square with the dot in the middle. The position of AT Cnc is marked by us with a blue star symbol. On the Chinese map it is clear that AT Cnc is close to another constellation (the Beacon Fire, highlighted with brownish line) but at the edge of the lunar mansion (white lines) of Yugui. On the Korean map the lunar mansion lines have the same position and are drawn thin and red. All constellations seem to be farther away from AT Cnc.}
  \label{fig:yuguiMaps}
  \includegraphics[width=\columnwidth]{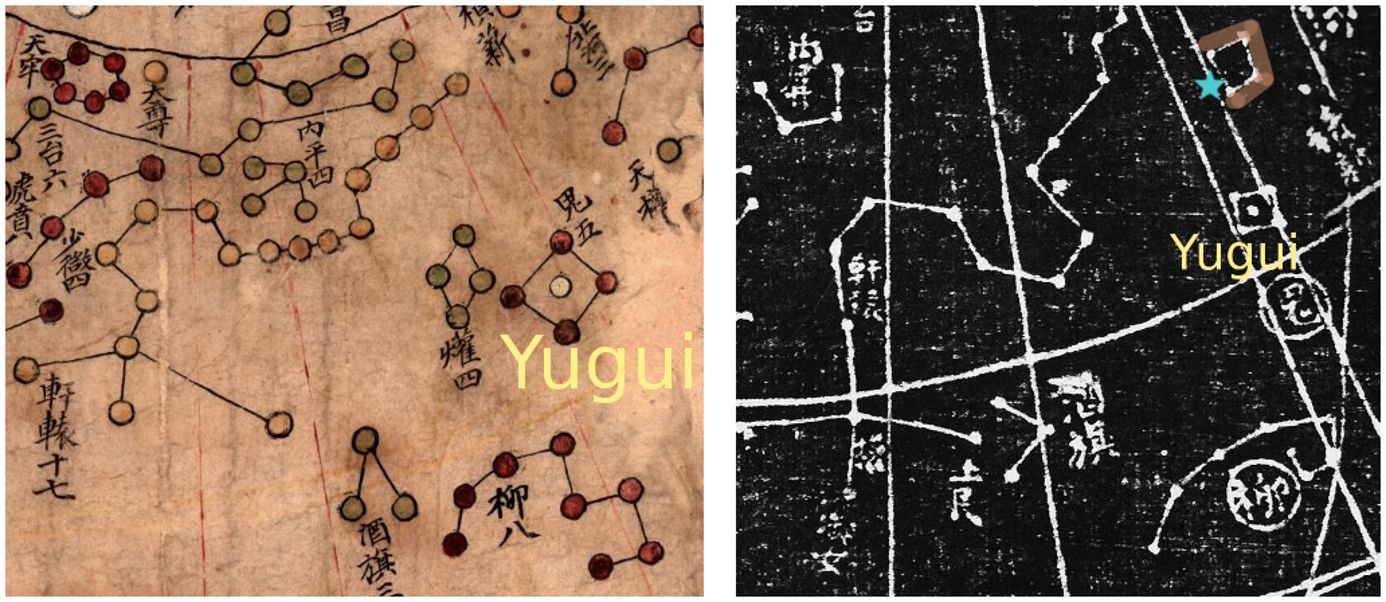} 
 \end{figure}

	The given date translates to February/ March 1645. Yugui (abbreviated Gui), The Ghost, is the quadrilateral of stars around the M44 star cluster. There is one other guest star in Yugui, namely on October 4th 1031 (also a Korean report) but this date does not fit the age of the shell. However, the fitting of the position 'in Yugui' to the position of AT Cnc is questionable: On the Chinese map, the asterism of the Beacon Fire is much closer than the asterism of the Ghost (see Fig.~\ref{fig:yuguiMaps}), so we would expect to describe an outburst of AT Cnc among constellations as 'trespassing against the Beacon Fire' which would have a different divinatory meaning and, thus, is maybe not relevant for the chronicle. On the Korean map it is not clear, if the asterism of the Beacon Fire is at the same place. 
% \onecolumn
% \twocolumn
	
	However, the distance of more than 4\degr\ of AT Cnc to the northern most star of Yugui is much too big for a likely coincidence -- especially in telescopic time. As also discussed by \citet{shara2017_ATcnc_steph} the term 'Yugui' also designates a lunar mansion and this lunar mansion could possibly include AT Cnc (the star is close to the border, see Fig.~\ref{fig:yuguiMaps}). However, astronomers usually use the term 'xiu' if they refer to the lunar mansion and this is not given in the text. That is why, we agree with the conclusion in \citet{shara2017_ATcnc_steph} that this record does not refer to the lunar mansion but to the constellation. Due to the large angular separation of the constellation and AT Cnc we disagree with their conclusion in \citet{shara2017_ATcnc_steph} that the event in '1645 CE is in intriguing proximity, both in time and location, to our deduced date of the last nova eruption of AT Cnc'. We consider the age as fitting but not the position.
	
	\subsection{Conclusion: No certain historical report for the AT Cnc shell}
	Hence, we agree that 'the evidence currently available is insufficient to support a claim of a likely coincidence' \citep{shara2017_ATcnc_steph} and not only `currently' but perhaps will never be. Recalling the genesis of the lists of guest stars (a huge number of exact astronomical observations filtered by astrological importance and, then, again filtered by chronicler's judgment of relevance, plus the random filter of the ravages of time) it is possible that the outburst of AT Cnc between 1552 and 1777 had been observed but the record is not preserved. Unfortunately, we cannot deduce anything from this. 
	
	\section{Case: Z Cam -- The origin of all historical considerations} 
	The whole discussion on usage of historical records for modern astrophysics seems to have started in 2007, when Shara et al. published an important paper in Nature \citep{shara2007} on the discovery of nova shell filaments around the dwarf nova prototypical star Z Cam. Since Z Cam itself is one of the closest Z Cam stars it is also one of the brightest ones (V-brightness between 14.5 and 10.0~mag) and the shell is extraordinary large; it covers almost the area of the full moon (radius $\sim15^\prime$, cf. \citet{shara2007}). It was detected in UV and X-ray by GALEX and imaged in H$\alpha$ by Michael Shara's team. The determination of the mass suits typical nova ejections and the trial to estimate the age of the shell ended up with a suggested age of 240 to 2400 years with a preference of the latter because of the observed snow-ploughing of the shell. Hence, we have to look for historical nova eruptions between $-393$ and $1767$ (with a preference of earlier dates). \citet{shara2007} state correctly ``Records of erupting stars two or more millennia ago are almost non-existent, so it is not surprising that no historical record of a Z Cam nova eruption exists.'' but two issues later, a reader of the magazine from Sweden suggests to consider the Chinese ``report of a `guest star' in October--November 77 bc (P. Y. Ho Vistas Astron. 5, 127--225; 1962)'' because ``The position in the sky fits Z Camelopardalis.'' \citep{johansson}.
	
	\subsection{The historical report} 
	Looking at the text witness, we find the text in three English translations: see Tab.~\ref{tab:event77}. At first glance, their different wording and spelling is irritating because the standard system of transliteration has changed: The current standard system is pinyin which is used by \citet{xu2000} but Ho and Hsi used another one. Their `Tzu-wei' equals pinyin `Ziwei' which describes the same asterism as the term `Zigong'. Respectively, the other terms in the various translations designate the same astronomical scene (see caption of Tab.~\ref{tab:event77}).   
	\begin{table}  
	\caption{Three translations of the historical record for Nova 77 BCE: Doushu is the modern transliteration of Tou-Shu, another name for Zigong is Ziwei (old spelling Tzu-Wei or Tzu-Wei-Yüan, the Purple Forbidden Enclosure), and the asterism of [Bei-]Ji is named the Northern Pole Asterism (not our Pole Star but a chain of $\gamma,\beta$ UMi, 4\,UMi, 5\,UMi, and HIP\,62572). Hence, the three translators interpret the same astronomical scenario, cf. map in Fig.~\ref{fig:zCamMap}.}
	\label{tab:event77}
	\begin{tabular} {|p{.27\columnwidth}|p{.27\columnwidth}|p{.27\columnwidth}| }\hline
	Xu et al. & Ho & Hsi \\ \hline
	Emperor Zhao of Han [China], 4th year of the Yuanfeng reign period, the 9th month. A guest star was situated within Zigong between Doushu and [Bei-]Ji.
	&  During the ninth month of the fourth year of the Yuan-Fêng reign period [of the Emperor Hsiao-Chao-Ti] a guest star appeared at the Tzu-Wei (Enclosure) between the star Tou-Shu ($\alpha$ UMa) and the Pole (star). 
	&  9th month, 4th year of Y\"uan-feng of Han dynasty, a guest star was found between the Pole Star and the Northern Dipper in the constellation Tzu-wei-y\"uan.
	\\ \hline
	Han shu, Tianwen zhi, ch. 26
	& CHS 26/29b; HHHY 28/3b; B(1); W38; L; Hsi
	& Han shu 
	\\ \hline
	\end{tabular} 
	\end{table} 

	The date translates to October--November 77~BCE ($= -76$). The four given asterisms designate three constellations and a single star: Zigong is the Enclosure of the Purple Palace, covering almost the whole circumpolar region. Doushu is a single star-asterism naming the star $\alpha$ UMa while Beiji is a group of five stars with $\beta$ UMi as the main star.\footnote{For identifications e.\,g. cf. \cite{sunKistemaker}, the appendix in \cite{xu2000}, and many popular summaries -- even the swarm intelligence wikipedia reflects this as common sense knowledge.} The Northern Pole Asterism (Bei Ji) contains five stars, i.\,e. the Emperor, the Crown Prince, the Concubine, her Son, and the Celestial Pivot. Obviously, this asterism at the centre of the Purple Forbidden Palace is the celestial representation of the most important players of the Chinese empire.

  \citet{ho} mentions that the object reported from 77 BCE had been suggested as nova not only by Hsi but also by Biot which is critical because the definition of novae changed in the 1930s\footnote{Three papers by F. Zwicky and W. Baade in 1934 and further developments until c. 1940 introduced the term `super-nova' for very bright cases of appearances and suggested stars made out of the newly discovered neutrons to be a result of supernova eruptions. Hence, they launched the research of the physics of supernovae as explosions of massive stars.} and again in the 1960s\footnote{Three papers by R. Kraft 1962 to 1964 suggested different physics behind the terms `novae' and `supernovae' with novae being eruptions on the surfaces of white dwarfs.}. Thus, the term 'nova' for us has a narrower meaning than for Biot (1846) and for Hsi (1957). For them 'nova' designates any dot-like object (star) that appeared for a limited duration (completely independent of its physical cause): It could be a Mira star, a flare star, a supernova, or any sort of outburst, e.\,g. an outburst of a cataclysmic variable which is \textit{the only} phenomenon \textit{we} would call a (classical) `nova'. Modern astrophysicists identify the term `nova' with an outburst of a CV (which, as if it was by chance) causes a brightening of the system and, therefore, might make an invisible system visible for a while. Hence, the notion of these early suggestions of the record to report a nova does not necessarily support the suggestion by \citet{johansson} which, thus, has to be discussed.

	\subsection{Discussion of the suggested identification(s)}
\begin{figure*}
   \caption{Map of the circumpolar region with Chinese asterism displays; epoch/equinox: $-76$ ($=$ 77~BCE). The asterisms mentioned in the suggested historical records (Tab.~\ref{tab:ZCam}) are highlighted in orange (walls of Zigong (palace), [Bei-]Ji and Dou-Shu). The yellow area is where we should look for the object that caused the guest star of 77 BCE, and Z Cam is marked in blue. Map produced with Stellarium Freeware and asterisms checked with the Suzhou Map. We appologize for the unreadably small lines in the upper left corner; they are an artefact of the genesis of the map and we did not delete them because users of the software will otherwise wonder what we changed in the map. The asterisms numbered in blue are Celestial Kitchen, Four Advisors, Celestial Bed, Royal Secretary, Maids-in-Waiting, Canopy of the Emperor, Six Jia (1\dots7 respectively). They are irrelevant for the nova discussion but serve to illustrate the celestial image of this culture.}
  \label{fig:zCamMap}
  \includegraphics[width=\textwidth]{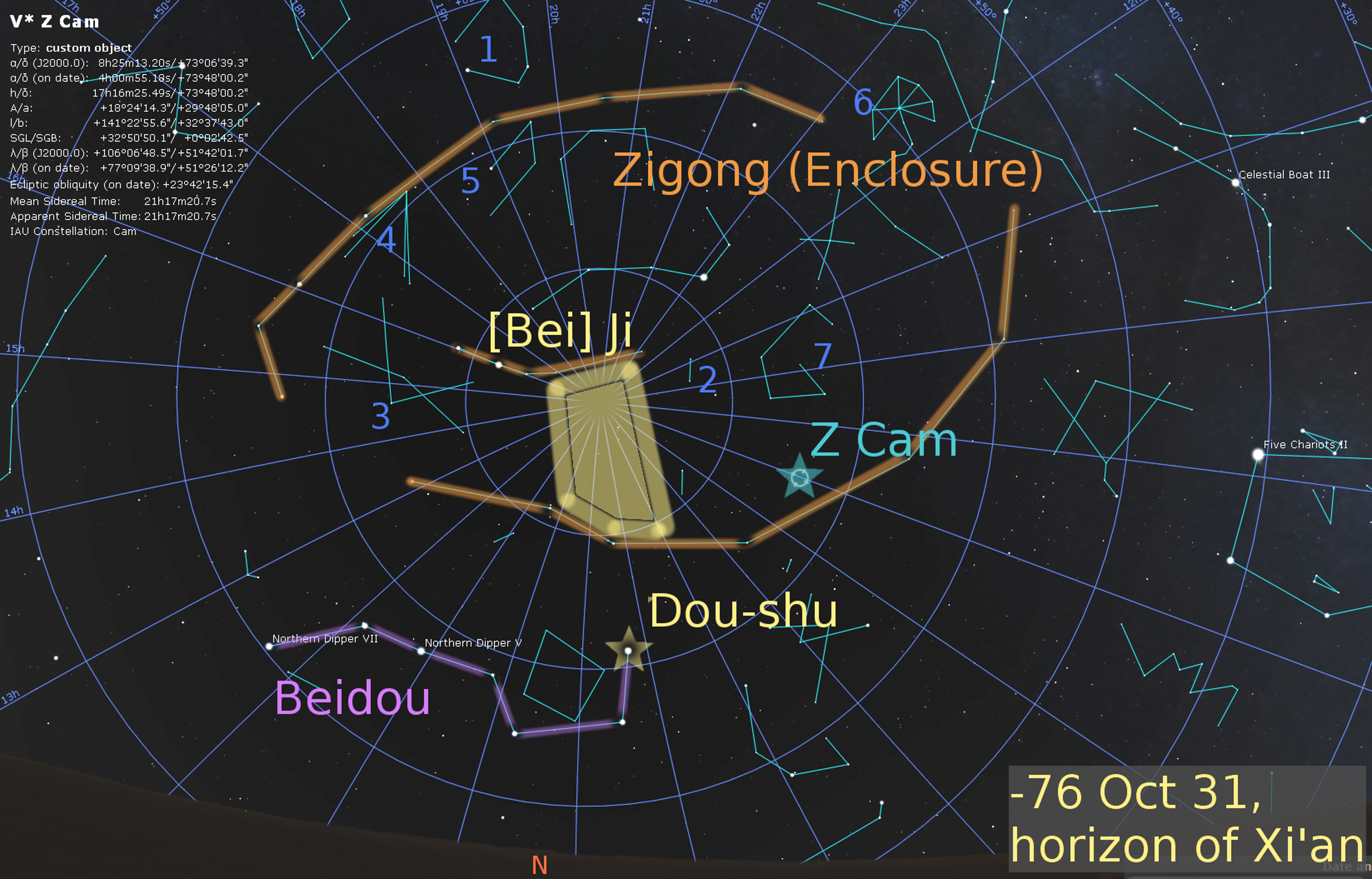} 
 \end{figure*}

 Consider the suggestion by G\"oran Johansson to identify this object with Z Cam: The position in the historical record is described as `within Zigong', the Central (or Purple) Palace, which locates the appearance within the circumpolar region. Fortunately, it is also described `between the [single-star asterism] Dou-shu and [the small asterism] Ji' and only a part of the area in between them is also within Zigong. The three asterisms and the resulting area to be considered are highlighted in our map Fig.~\ref{fig:zCamMap}. However, the distance between Z Cam and the area described in the text is more than 10\degr . This is too big for naked eye observers (20 diametres of the full moon) as well as in a historical epoch where the accuracy for astrometrical measurements had been in the range of 1\degr\ or better (cf. \citet{needham} and \citet{sunKistemaker}). Additionally, the divinatory meaning of the Zigong palace is the Emperor's place in the Forbidden City: It would make a huge difference for the intention of both, the astrologer as well as the chronicler, if the guest star appears next to the Emperor himself or his sons (asterism Beiji) or somewhere else close to the eastern wall of Zigong and closer to the asterism of the Celestial Kitchen (in Draco), the Four Advisors or the Prime Minister (in CVn, close to UMa).    
%\onecolumn 

 \begin{table*}
  \caption{There are several suggestions for historical records of appearances which are proposed to have caused the Z Cam shell. The table displays the suggested records by Johannson and Warner which all fit the age of the shell as determined by \citet{shara2007}.} 
  \label{tab:ZCam}
  \begin{tabular}{p{.21\textwidth}p{.21\textwidth}p{.21\textwidth}p{.21\textwidth}}
   suggested record & author of suggestion & text of the record according to \citet{xu2000} & where?\\
   \hline
 Chinese guest star 77 BCE
  & \citet{johansson} in reaction to \citet{shara2007}
  & Emperor Zhao of Han [China],
4th year of the Yuanfeng reign period, the 9th month.
A guest star was situated within Zigong between Doushu
and [Bei-]Ji.  
	& area highlighted in yellow in Fig.~\ref{fig:zCamMap}\\ 

 Korean guest star 158 CE
 & \citep{warner}
  &13th year of King Chadae of Koguryo [Korea], 10th month.
There was a `fuzzy star' in Beidou. 
 & within the Big Dipper \\ 
 Chinese guest star 305 CE
 & \citep{warner}
 & 2nd year of the Yongxing reign period, 7th month, day
dingchou [14]. There was a `fuzzy star' in Beidou.
	& within the Big Dipper\\ 
 & &	\multicolumn{2}{p{.5\textwidth}}{ }\\
 note: 	&	\multicolumn{3}{p{.73\textwidth}}{(1) The translation `fuzzy star' of the Chinese `xing bo' is a common alternative to `bushy star' or `rayed star'. Hsi usually translates `sparkeling star'.}\\
		&	\multicolumn{3}{p{.73\textwidth}}{(2) Concerning the wording in general: If something is described as \textit{fuzzy} this does not mean that it was a comet: A humid or dusty atmosphere can have the same effect on a shiny bright star-like object. Some bright historical supernovae records report the object fuzzy (e.\,g. Japanese record from Ichidai y\=oki for the SN~1054, \citet[p.\,139]{xu2000}). As long as there is no tail (with length and direction) reported, we cannot exclude such reports as candidates for novae and supernovae.} \\
 \end{tabular} 
 \end{table*} 
% \twocolumn
	
	\citet{warner} alternatively suggested `identifications with novae near $\alpha$ UMi in AD 158 and AD 305'. This is definitely wrong, because of several misunderstandings: 
	\begin{enumerate} 
	 \item Warner cites Nickiforov's nova list \citep{nickiforov}. Dating 2010, this is the most recent list -- but it is full of mistakes, e.\,g. 
	  \begin{itemize} 
	   \item Nickiforov did not understand the astronomical counting of years (with `minus' and a year `0' which does not exist in the `common era'-counting). Therefore, all his dates `BCE' are systematically shifted by one year. 
		\item Nickiforov always gives single stars, even if the text does not. In the cases cited by Warner, the texts speak of constellation Beidou (Northern Dipper) which is identical with the Big Dipper. It is an asterism covering an area of roughly $10\degr\times30\degr$ and not a single star.
		\item In this special case, Nickiforov even gives the wrong star: The main star of Beidou is $\alpha$ UMa and Nickiforov gives (incorrectly) $\alpha$ UMi ($\geq30\degr$ away from the Big Dipper). 
	  \end{itemize} 
	 \item Warner copied Nickiforov's mistake and even stresses it: Warner took $\alpha$ UMi for the star close to Z Cam which is not the case. The distance is roughly 17\degr . 
	 \item The correct main star of Warner's suggested guest star asterisms in 158 and 305 CE is $\alpha$ UMa. This star is indeed (accidentally) the star of the constellation of Beidou which is closest to Z Cam. Still, its distance from Z Cam is roughly 18\degr\  and, thus, too far away for a likely match.
	\end{enumerate}
	
	To sum it up, all the distances of the locations given in the historical reports from $+158$ and $+305$ are much too far from the position of the object in the sky (Z Cam). Additionally, Beidou is an asterism outside the celestial enclosure of the Purple Palace (Zigong), while the area suggested by the record in 77 BCE and Z Cam are inside this enclosure. That means, the divinatory meaning of an appearance in the Beidou (Dipper) would be completely different and ancient Chinese astronomers would certainly not have mixed that up. Thinking from the perspective of the historical authors of these astronomical records we, therefore, have to conclude that any suggestions of guest stars in asterisms outside the Zigong enclosure (even if they have a comparable distance to Z Cam) are highly unlikely to have caused the shell of Z Cam. Hence, we cannot support any of the above suggestions. 
	
	\subsection{Conclusion} 
	There is no historical record that fits the position of Z Cam. If Oriental astronomers observed this star outbursting during the last 2000 years, their observation might be unreported or lost. 
	
 Which information can we derive from historical reports? For the central question of this article, we have to resign in the case of Z Cam because (as already suggested by \citet{shara2007}) no historical counterpart can be identified.
 
  \section{Case: Te 11 as Nova 483 Ori}
    In our introduction, we also cited the example of Te~11 with which \citet{miszalski2016} wanted to derive conclusions on the time-scale of the evolution of the system. They admit that the time-scale of a few hundred years between recurrent nova explosions `is poorly constrained for Te~11, although the $\sim1500$~yr time-scale from last recorded nova eruption to the present state of regular DN (disc instability) outbursts in Te~11 is consistent with the evolution/variation of mass transfer rate in short-period CVs through a nova cycle (e.\,g. BK Lyn; Patterson et al. 2013).' The reference to \citet{pat2013} is a weak support because we have proven the case of BK Lyn to be likely but not certain. 

 \subsection{The historical record}
 The identification of the object Te~11 (initially listed in catalogs of planetary nebulae until it was re-interpreted as a nova shell) with the guest star observed in China in November/ December 483 CE has been suggested since \citep{warner1995} but the historical record is rather vague: Tab.~\ref{tab:event483}. 
 	\begin{table}  
	\caption{Two translations of the historical record of the transient in 483 CE.}
	\label{tab:event483}
	\begin{tabular} {|p{.37\columnwidth}|p{.37\columnwidth}|p{.07\columnwidth}| }\hline
	Xu et al. & Ho & Hsi \\ \hline
	Emperor Xiaowen of Wei [China], 7th year of the Taihe reign period, the 10th month. There was a guest star east of Shen, as large as a peck measure and like a fuzzy star.
	& During the tenth month of the seventh year of the Thai Ho reign-period a `guest star' of the size of a peck measure appeared at the \textit{Shen} (21st lunar mansion). It looked like a \textit{po}
	&  --  
	\\ \hline
	Wei shu, Tianxiang zhi, ch. 105
	& WS 105/3/37a
	&  --
	\\ \hline
	\end{tabular} 
	\end{table} 

  Hsi excluded this transient because of the terminology but as it has been explained many times since then, the fact that something is described as `bo xing $=$ po hsing' or \textit{fuzzy} does not exclude it as a comet: A bright transient observed through fog, dust, or smog might appear fuzzy (cf. Tab.~\ref{tab:ZCam}) or observed with astigmatism and spherical aberration might also appear to have rays (cf. \citet[p.\,44]{steph77}). Even transients that are definitely identified as supernovae are sometimes reported as fuzzy (cf. records from Japan of SN~1054, \citet[p.\,138/9]{xu2000}, from China of SN~1572, \citet[p.\,143]{xu2000}).

\begin{figure}
   \caption{Map of the Chinese asterism Shen; left on the historical Suzhou map (black marble), right in Stellarium with the object Te~11 marked. We also sketched the area `east of Shen' which is the position described by the historical record.}
  \label{fig:Te11Map}
  \includegraphics[width=\columnwidth]{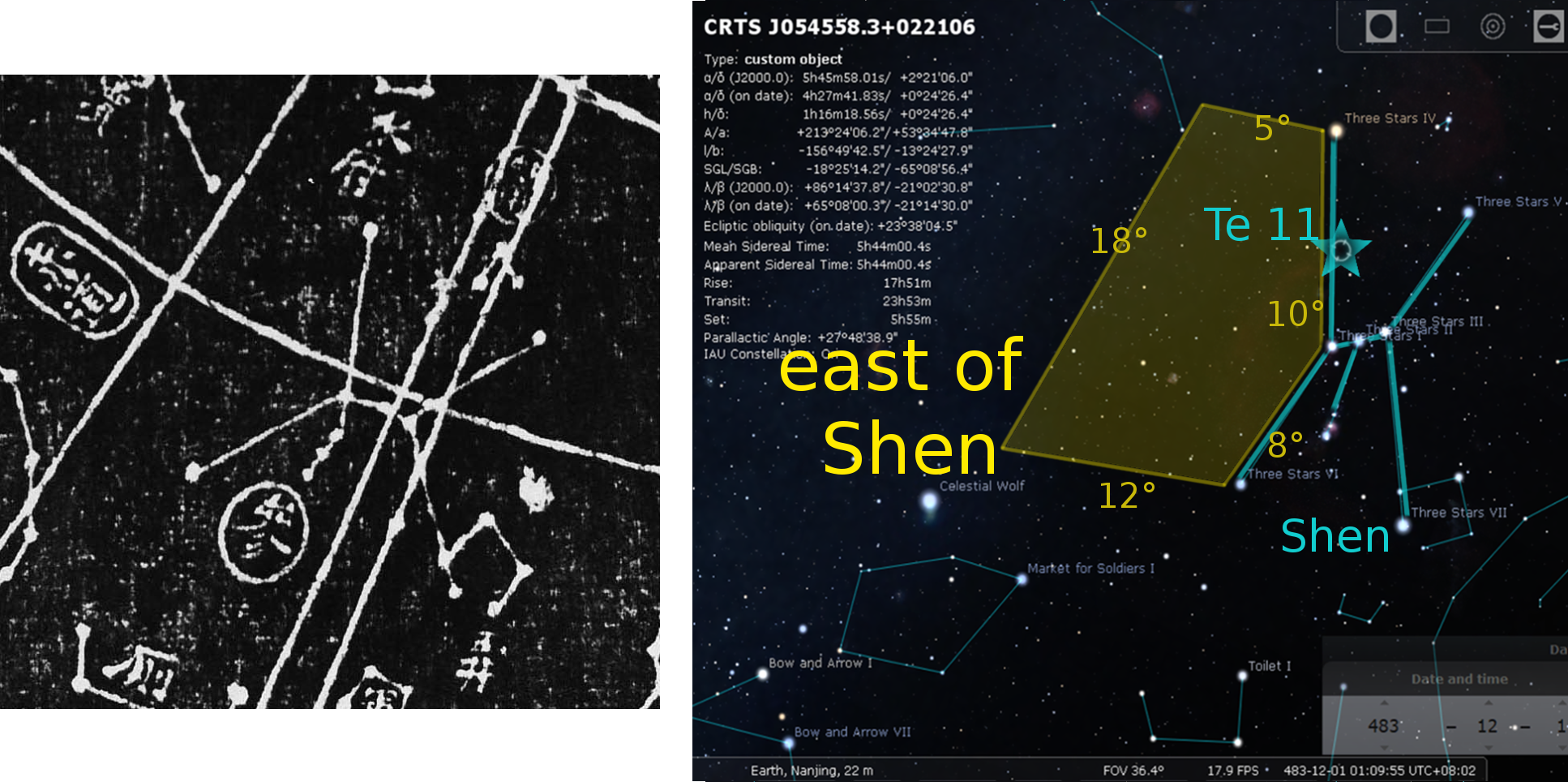} 
 \end{figure}

 The position is given as `east of Shen' where `Shen' (translated: Three Stars) consists of the seven bright stars of our constellation of Orion. The position `east of Orion' implies a huge field, a pentagon with sites of roughly $8\degr, 10\degr, 5\degr, 18\degr, 12\degr$ (see Fig.~\ref{fig:Te11Map}). We admit that Te~11 is at the eastern edge of the asterism of Shen, roughly 5\degr\ north of $\zeta$~Ori which makes it a possible identification. However, a rigorous reading of the text would locate the guest star east of Te~11 (see Fig.~\ref{fig:Te11Map}). 
 
	\subsection{Conclusion} 
	We appreciate the idea of cross-checking existing transient surveys with newly discovered PNe \citep[p.\,639]{miszalski2016} and support the new interpretation that Te~11 might be a remnant of an ancient nova. The identification of Te~11 with the Chinese guest star in 483~CE we consider as possible but weak and having alternatives (as in the case of BK Lyn). 
 
 \section{Case of the potential Nova -47} 
\begin{figure*} 
  \caption{Map for locating the historical record, all CVs in the field (green dots: CVs fainter than 18~mag, red dots: brighter than 18~mag). This map was generated by the author $\sim 1$~year before the appearance of the paper by \citet{goettgens2019}. The coordinates suggested by $Hsi$ seam to be wrong; those by $Xi+$ were centered at $\lambda$~Sgr at an epoch/equinox 1950 -- this explains the offset towards west but centering the star was a misreading anyway. In the field defined by Stephenson and us, there are $\sim20$ stars which fit the reported position (more or less). The best fitting stars are marked with a (red) star symbol: To the equinox of the record the star "`Establishment III"' was almost directly east of $\lambda$~Sgr; due to precession Establishment and, therefore, the CV candidates are now northeast of $\lambda$ Sgr. The dots in the left half of our circle are formally east of the mentioned star but less likely with regard to the standards of interpreting the historical text. The globular cluster M22 fits the position within the historical margin of error.}
  \label{fig:event-47}
  \includegraphics[width=\textwidth]{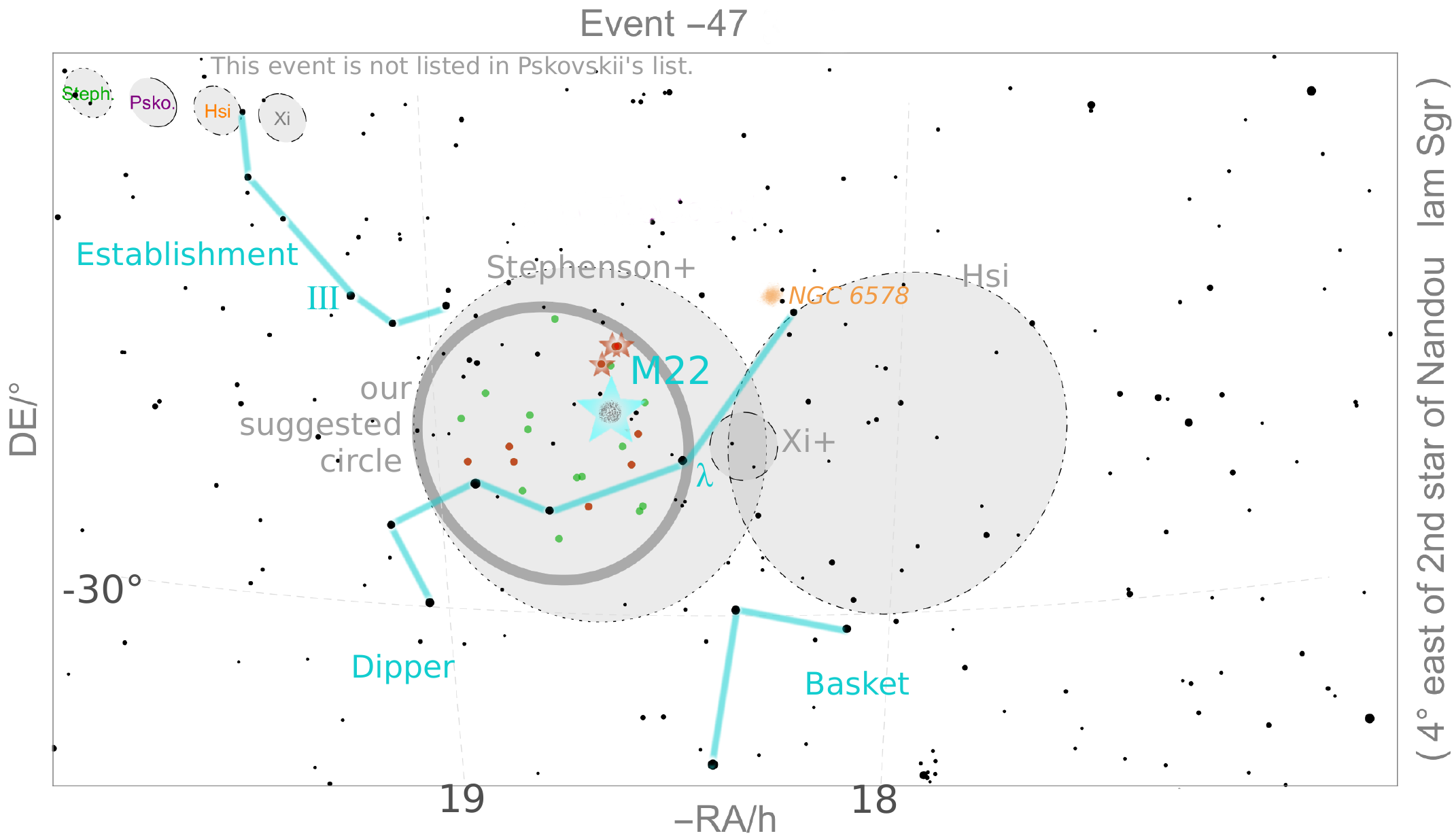} 
 \end{figure*}

 Finally, another suggestion for an old nova appeared during the referee process: \citet{goettgens2019} `detected a small emission nebula in NGC~6656 at a distance of about $14\arcsec$ from the cluster centre'. The globular cluster NGC~6656 ($=$Messier 22) in which this old nova remnant was found indeed fits roughly the position `east of the second star of Nandou'. The text of the record reports it `about four chi' ($~4\degr$) away from this star (cf. \citet[p.\,130]{xu2000}) while the cluster is only $\sim2\fdg5$ east of the second star of Nandou ($\lambda$ Sgr): see Fig.~\ref{fig:event-47}. 
 	\begin{table}  
	\caption{Three translations of the historical record for the guest star in $-47$: All giving the same position, i.\,e. $\sim4\degr$ east of the second star of the Southern Dipper (in Sgr). The mentioning of [LM8] in Xu's text only makes sure that the asterism name designates the determining asterism for a certain lunar mansion (LM). It does not imply that Nandou, in this case, refers to the LM (which would not make sense because there is no 2nd star of any LM).}
	\label{tab:event47}
	\begin{tabular} {|p{.27\columnwidth}|p{.27\columnwidth}|p{.27\columnwidth}| }\hline
	Xu et al. & Ho & Hsi \\ \hline
	Emperor Yuan of Han, 1st year of the Chuyuan reign period, the 4th month. There was a guest star as large as a melon with a bluish-white color about four chi east of the second star of Nandou [LM 8]. 
	& a guest star of the size of a melon and with a bluish-white colour, was seen about 4 chhih away east of the second star in the Nan-Tou.
	& 4th month, 1st year of Ch'u-yüan of Han, star as large as a melon, blue-white 4 chh'ih east of No. 2 of Nan-tou.   
	\\ \hline
	Han shu, Tianwen zhi, ch. 26
	& CHS, WHTK, HHHY, B, W, L, Hsi
	& Han shu 
	\\ \hline
	\end{tabular} 
	\end{table} 
  \begin{figure} 
  \caption{Chinese Suzhou map (white on black marble) with highlighted position of M22 and star numbers. It clearly shows that they considered the Milky Way (coloured by us) to extend to the position of the cluster.}
  \label{fig:nandou}
  \includegraphics[width=\columnwidth]{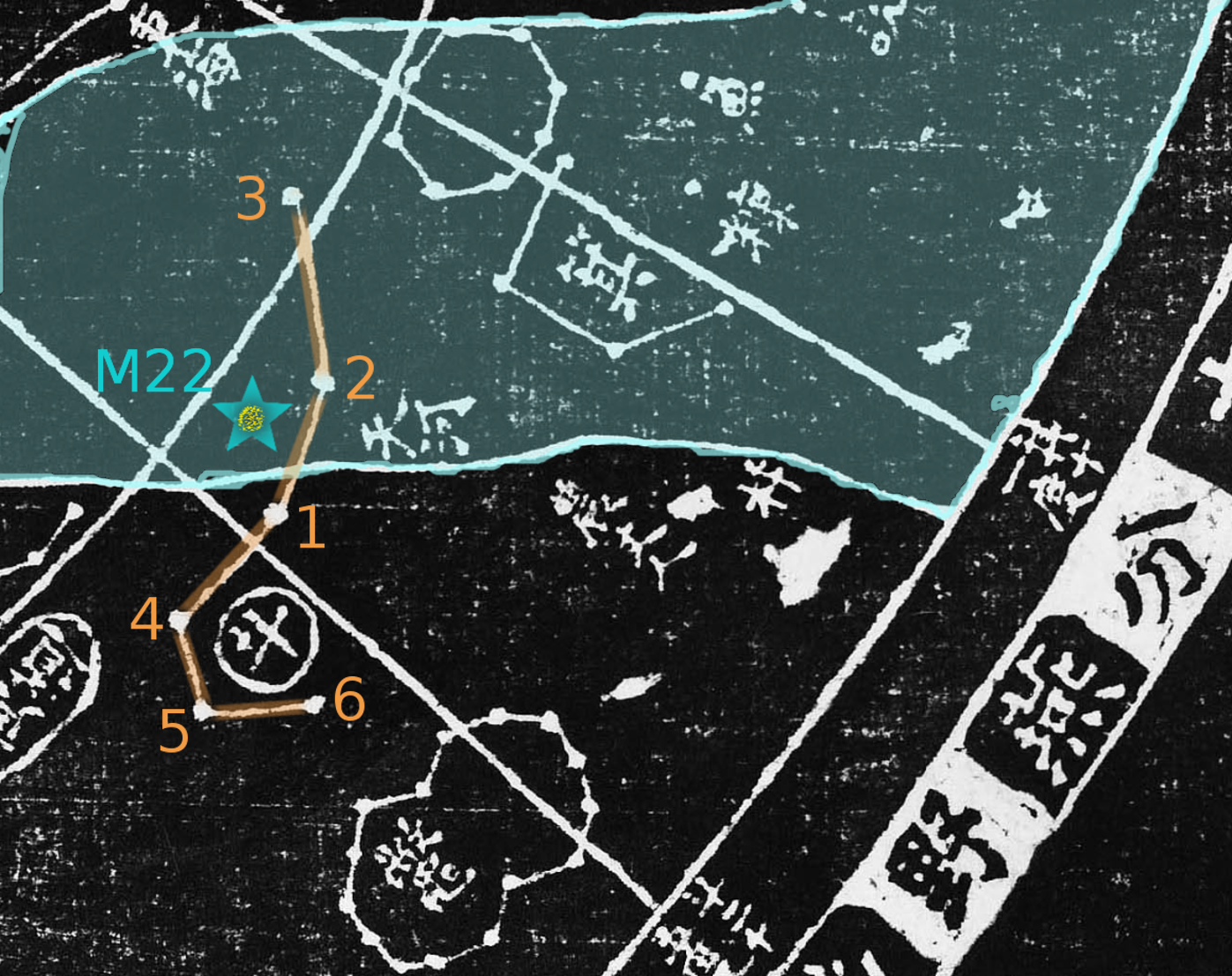} 
 \end{figure}
  	\begin{table}  
	\caption{Suggested coordinates plotted in Fig.~\ref{fig:event-47}, all (RA,DE)$_\textrm{1950}$.}
	\label{tab:coord47}
	\begin{tabular} {|p{.27\columnwidth}|p{.27\columnwidth}|p{.27\columnwidth}| }\hline
	\citet{hsi} & \citet{xi+po} & \citet{steph77} \\ \hline
	 (18\,h, $-25\degr)$
	&  (18\,h\ 20\,m, $-25\degr)$
	&  (18\,h\ 40\,m, $-25\degr)$  
	\\ \hline
	\end{tabular} 
	\end{table} 
	
 This report is one of the rare cases with startling clear positioning: The chain of Nandou is identified with $\mu,\lambda,\phi,\rho,\sigma,\tau,\zeta$ Sgr (from north to south), the second star is $\lambda$~Sgr -- no matter if we count from north to south or from the main star ($\phi$~Sgr) towards north. Yet, the positions suggested by the authors of the 20th century are rather divers (Tab.~\ref{tab:coord47}). Hsi suggests the position as `near NGC~6578' but this planetary nebula (PN G010.8-01.8) is closer to the third star of Nandou ($\mu$ Sgr) than to the second. Nandou is one of the asterisms in which the numbering of stars starts in the middle of the chain because $\phi$ Sgr is closest to the right ascension line which separates two lunar mansions from each other (cf. Suzhou map in Fig.~\ref{fig:nandou}).	
 
 In Fig.~\ref{fig:event-47} the stars marked with a little red $\star$-symbol are V4021 Sgr, OGLE-BLG-DN-1057, and OGLE-BLG-DN-1056. Their position roughly fits the historically given position -- but could they really have caused the event in $-47$? V4021~Sgr showed a classical nova outburst with an amplitude of $\sim9$~mag in 1977. If this was the recurrent outburst of something ancient, the ancient one would also have been below the detection limit for the human eye. The two dwarf novae of the OGLE survey have to be studied in more detail -- as well as all the other CVs in the above field. 
 
 However, one might discuss the identification of the newly discovered nebula in M22 with the guest star with regard to the determined age of the shell and visibility for human observers. In their chapter `Nova and nebula brightness', \citet{goettgens2019} compute (distance modulus) the apparent brightness of a nova with typical absolute magnitude of $-7\pm1.4$~mag at a distance of $3.2$~kpc to the order of $\sim5.5\pm1.4$~mag (neglecting the error bar of the distance of the cluster and the known extinction to M22 by ISM in between). From this, they conclude that `40\,\% of all novae in NGC~6656 reach an apparent brightness of 5~mag or brighter, which could be seen with the naked eye.' We doubt that this argument is valid because the globular cluster itself has a brightness of 5.1~mag and any nova inside it should be much brighter than the background brightness of the cluster. Additionally, there are several arguments against the detection limit of 5~mag for noticing transients by the naked eye observers (cf. \citet{vogt2019}). 

\begin{figure*} 
  \caption{A map of Ptolemy's stars in front of the average background brightness of the area. The colours indicate whether the brightness given in the old catalogue is too bright (green colours) or too faint (red colours) or correct (blue); graphics by Philipp Protte. Similar plots for the star catalogues by as-Sufi and Brahe show roughly the same distribution. The magnified region in the circle is the Chinese Southern Dipper (Nandou).}
  \label{fig:ptolMag}
  \includegraphics[width=\textwidth]{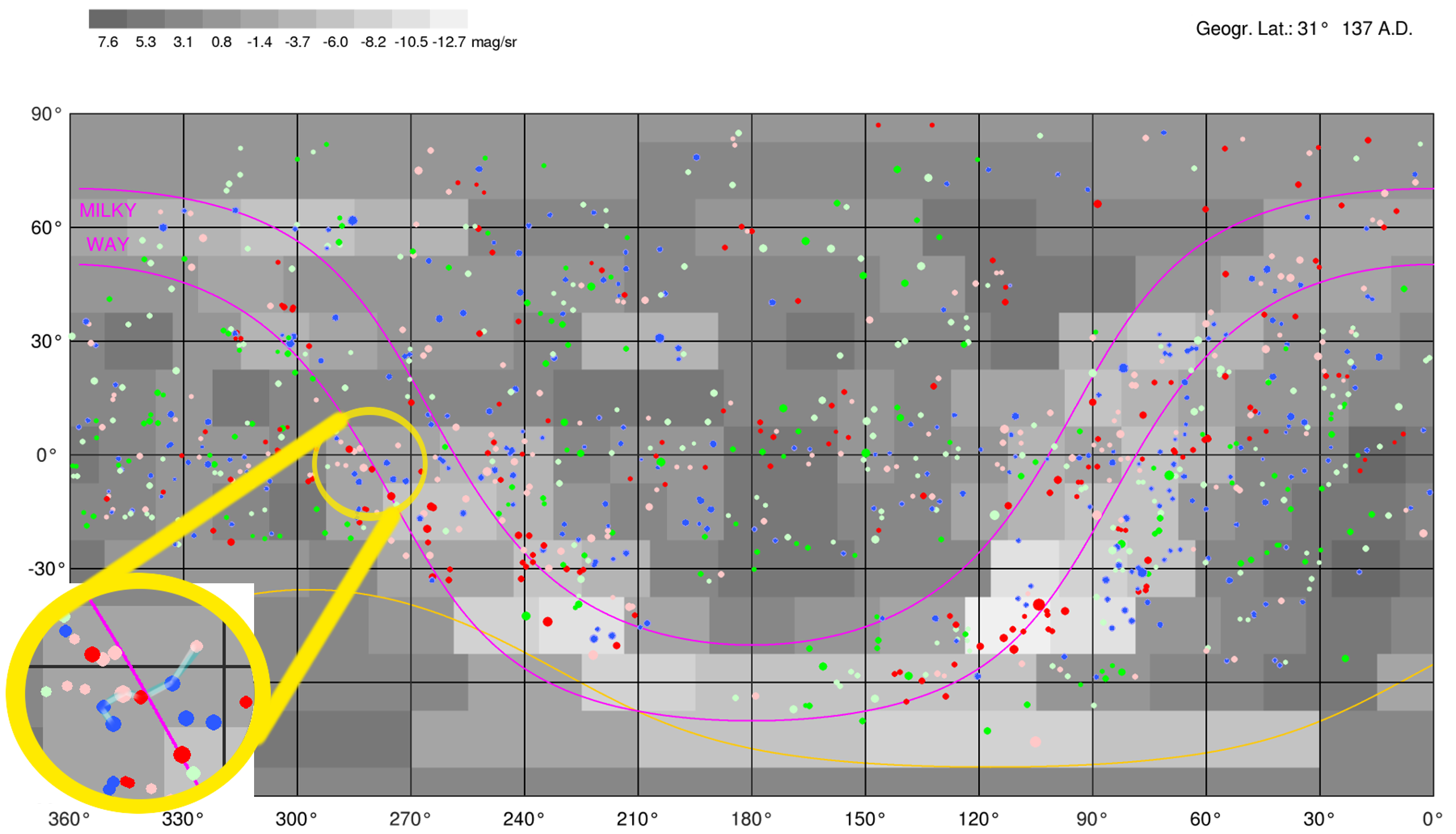} 
 \end{figure*}

 Further, \citet{goettgens2019} estimate `About $4-5$\,\% of all novae reach an apparent brightness of at least 3~mag' and use this as an argument that a nova inside the cluster `could have been visible to Chinese observers'. We admit, that this chance is not zero but still smaller than $4-5$\,\% because of the proximity of the galactic centre with its bright clouds of the Milky Way and many bright stars (Fig.~\ref{fig:ptolMag}). In a vicinity of bright stars as well as against bright background all old star catalogues (such as Ptolemy's, as-Sufi's, and Brahe's which provide own measurements of magnitudes) estimate stellar brightnesses too faint (see Fig.~\ref{fig:ptolMag}). That means, that a faint target like the globular cluster M22 was neglected in those catalogues. These catalogues are almost complete only for the stars brighter than $4$~mag (cf. for Ptolemy's catalogue: \citet[p.\,247]{smh2017}). M22 in particular is at the edge of the Milky Way and could even be considered as a brighter knot of its clouds. This requires a high brightness for any transient in this region to be realized by humans -- even by trained and professional astronomers who are certainly able to discover newly appearing stars of even 4~mag (but unlikely fainter) in areas of the IAU-constellations Hercules or Pisces with dark background and few/no bright stars. These facts make the visibility of a nova in M22 still possible but unlikely for an ancient human observer.
 
    Additionally, \citet{goettgens2019} derive an age of the nebula of $2.0^{+0.8}_{-0.5}\cdot10^3$~yr from the nova remnant dimming rate ($10\pm3$~mmag/yr, \cite{duerbeck1992}) and the nova brightness distribution. This age would fit the observational range of Oriental guest stars. However, the calculation appears suspicious: \citet{goettgens2019} did not identify any CV which could have caused the nebula; they even discuss the possibility of the nebula being a PN and, therefore, caused by a solar mass giant star instead of a runaway thermonuclear explosion on the surface of a white dwarf. As they do not know a binary system having caused the nebula they cannot use its (current) brightness for their further estimates. Thus, the input data for this computation appears questionable and incomplete because $(i)$ the current brightness measured by \citet{goettgens2019} is the brightness of several spectral lines and not the brightness of the V-band continuum which should be the input parameter according to \citet{duerbeck1992}, $(ii)$ because the used dimming rate in the original paper \citep{duerbeck1992} did not concern the nova shell but the star on which the eruption took place. 

 Due to absent information concerning the input parameters for Duerbeck-equation, we would like to consider the expansion velocity: As the nebula appears as an ellipse of $2\farcs5\times2\arcsec$ \citet{goettgens2019} give the corresponding size at a distance of 3.2~kpc as 0.04~pc $\times$ 0.03~pc. The radius is, therefore, $\sim0.018$~pc. If the observed cloud expanded to this radius within 2064 years (from the guest star seen in $-47$ to the observation with MUSE in $2015-2017$) the average expansion velocity would be 8.5~km\,s$^{-1}$ which contradicts typical nova shell expansions in the order of some hundred or thousand km\,s$^{-1}$ (cf. \citet{valle1997}: FH Ser 1970 line of sight(LOS) velocity from 420~km\,s$^{-1}$ at the accretion disk to $\sim500$~km\,s$^{-1}$ at outer parts, whilst QU Vul 1984 average LOS velocity $1190\pm105$~km\,s$^{-1}$). A velocity of $\sim8.5$~km\,s$^{-1}$ is even smaller than typical expansion rates of PN (42~km\,s$^{-1}$ according to \citet{jacob2013}). Within a globular cluster there is almost none ISM to decelerate the expansion. That means, in any of the suggested interpretations, the velocity is too low. Since the size is properly measured, the time (age) should be smaller: In case, the object is a PN it's age should be $~400$~years and in case, it is a shell of a classical nova the age is only some decades. Thus, an observation of an eruption should have been made (if at all) in telescopic time.  

 \subsection{Conclusion} 
 Although the position of the historical Chinese guest star in $-47$ ($=$48~BCE) roughly fits the position of the nebula discovered in the M22 globula cluster, we doubt that this nebula could be old enough to be a remnant of an eruption roughly 2000~years ago. 

 We curiously await further observations of this target and the follow-up papers by Göttgens+ (which seem to be already in preparation) but suggest to drop the connection to this very ancient Chinese guest star.  
 
\section{Summary and Conclusions}
	First, we have to conclude that the suggested and astrophysically treated identifications are not reliable or proven to be wrong: 
 \begin{itemize} 
	\item The identification of BK Lyn with the Chinese guest star reported in 101~CE is possible (and likely) but has alternatives because there is more than one way of counting the stars within the Chinese asterism of Xuanyuan. 
	\item The Korean guest star in 1437 in Sco does not match the position of the nova shell found by \cite{shara2017_nov1437}. At the given position of the event in 1437 there are two CVs both not being able to have caused a brightening in the observed way. If the eruption of the star considered by \cite{shara2017_nov1437} was observed by any humans at all, the record might be lost or non-extant. 
	\item For the age-dated shell of AT Cnc ( $\sim330^{+135}_{-90}$ yr) there is no adequate guest star record in the lists of possible Oriental novae. 
	\item None of the suggested historical observations (in the years 77 BCE or 305 CE in China or 158 CE in Korea) fits the position of Z Cam. They are all $\sim15\dots20\degr$ away from the star and its shell. 
	\item The identification of Te~11 with the guest star of 483~CE is not completely excluded but has alternatives. 
	\item The nebula in M22 almost fits the historical position of a guest star in $-47$ but we doubt that the age of this shell fits this historical date. 
 \end{itemize}  

 \begin{figure} 
  \caption{Stemma for the modern lists of novae among Far Eastern guest star observations (a stemma is the graphical representation of text history). \textbf{Colours:} gray $=$ lists of point coordinates, blue: collections of text records, red: development of `nova' term.}
  \label{fig:filters}
  \includegraphics[width=\columnwidth]{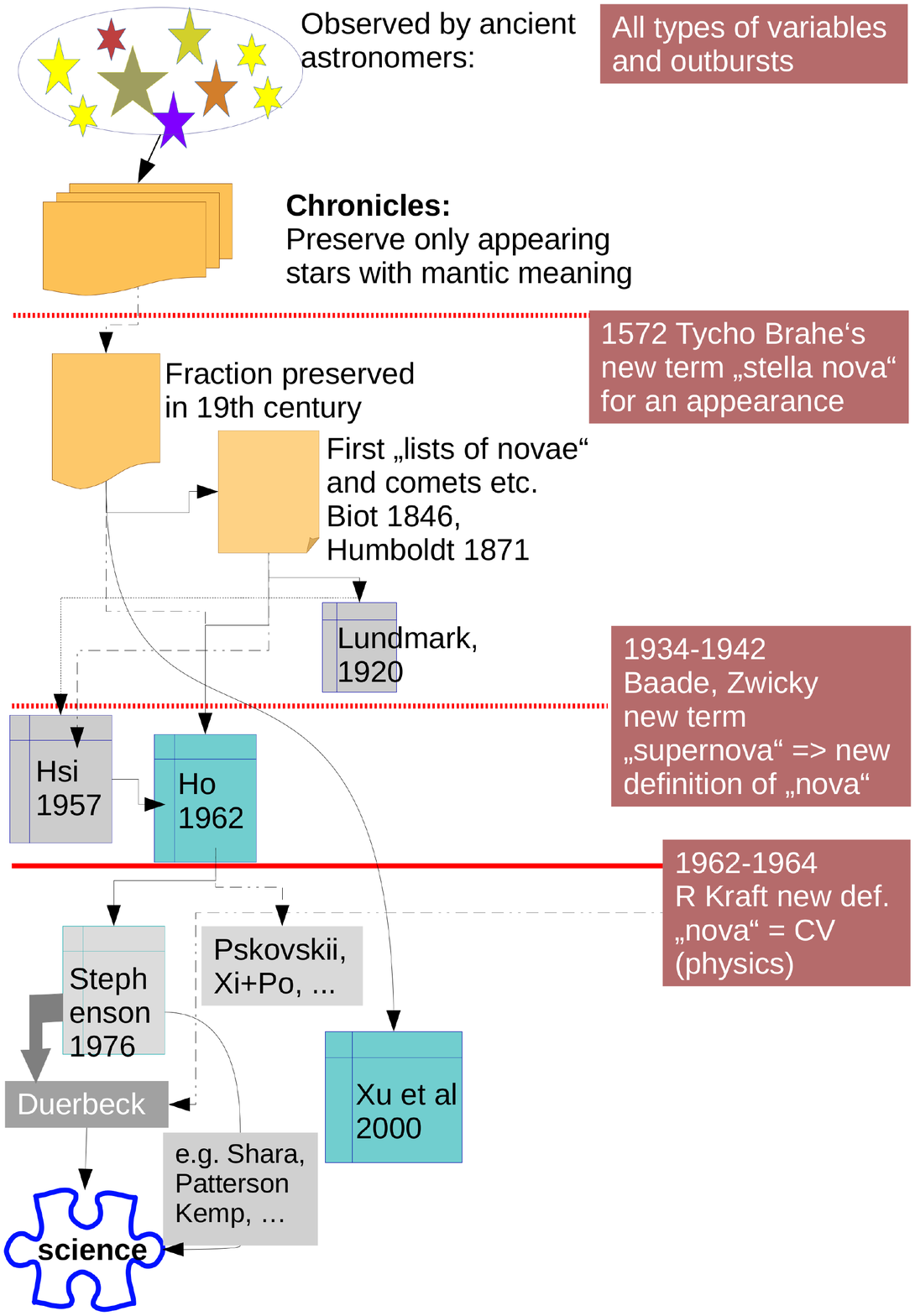} 
 \end{figure} 
 
 Second, it is difficult to draw conclusions concerning the evolution of binaries on the base of historical records: The genesis of the lists of historical Far Eastern novae (see stemma in Fig.~\ref{fig:filters}) shows that all our modern lists of potential nova records are incomplete by default. Most of the real records of astronomical observations are lost and only those records remain which had significance for any divination and politics. Of course, non-extant observational records of an object during centuries or millennia do not allow any conclusion concerning the outburst behavior or mass transfer rates or hibernation -- it only leads to the conclusion that no observational records are \textit{preserved} (not even that they did not exist). Additionally, CVs have typical brightnesses fainter than 12~mag. Dwarf nova outburst typically show amplitudes of a few (2--6) mag, while classical novae get much brighter (typically $11$ to $13$~mag but even $16$~mag are possible). That means, classical novae may have been visible to the naked eye but historical dwarf novae may not. Hence, even if naked-eye astronomers would have monitored an object after a classical nova systematically (which they certainly did not do), they very likely did not recognize dwarf nova outbursts some decades or centuries later. These facts support the conclusion that historical records unfortunately cannot contribute much to the questions of binary evolution. 

 In summary, the astrophysics can lead to questions on the historical transmission but the uncertainties of historical records are rather big for deriving information for astrophysical models such as the time scale of a post-nova-to-dwarf nova metamorphosis. We recommend caution in using historical texts and data for modern physics and suggest focusing on a conservative physical modelling. A non-matching of these models with historical observation or an apparent non-existence of historical observation could only demonstrate a lack of historical monitoring and/or gaps in the preservation of data (cf. Fig.~\ref{fig:filters}). It does not necessarily support or contradict the physical model. The contribution of historical data, in most cases, turns out to be a nice addition but should not be constraining.

\section*{Acknowledgements}
 The author thanks Claus Tappert (U Valparaíso, Chile) and Nalini Kirk (Excellence Cluster Topoi, Berlin, Germany) for a fruitful discussion of the astrophysical and historical Chinese content, respectively, Philipp Protte (U Jena, Germany) for his analysis of magnitudes in the ancient star catalogues in his master thesis, and Nikolaus Vogt (U Valparaíso, Chile), and Jesse Chapman (University of California, Berkeley) for improving the readability of the long paper. Ralph Neuh\"auser (AIU, U Jena, Germany) had the initiative and idea to reconsider historical nova identifications including new nova candidates. However, he did not agree with parts of this paper. We thank the State of Thuringia for financing the project. I additionally thank the anonymous referee for the strong interest in the topic, and excellent comments that improved my presentation of the transdisciplinary work tremendously.
 
\section*{Digital Maps} 
 \begin{description} 
	\item[Chinese Suzhou] \url{http://www.chinesehsc.org/zoomify/suzhou_planisphere.html} 
	\item[Korean Cheonsang Yeolcha Bunyajido] \url{https://digicult2.thulb.uni-jena.de/digicult/rsc/viewer/digicult_derivate_00115090/Sternkarte2.tif} 
	\item[Stellarium] \url{http://stellarium.org}, latest version used is 0.17.0, because 0.18.$\ast$ versions include more modern Chinese constellations and are, therefore, not wrong but a little bit confusing for this purpose. The releases 0.19.$\ast$\ are very useful again; they provide three Chinese sky cultures, namely the traditional (Suzhou) as `Chinese', the `Chinese Contemporary' which are the IAU-constellations combined with Chinese star names, and `Chinese Medieval' (based on the observational data of Huanyou, 1052 CE, Song dynasty) contributed by Sun Shuwei. 
 \end{description}	

%%%%%%%%%%%%%%%%%%%%%%%%%%%%%%%%%%%%%%%%%%%%%%%%%%


%\nocite{*}
\bibliographystyle{mnras}
\bibliography{chinGuestStars} % if your bibtex file is called example.bib

\begin{thebibliography}{}
\makeatletter
\relax
\def\mn@urlcharsother{\let\do\@makeother \do\$\do\&\do\#\do\^\do\_\do\%\do\~}
\def\mn@doi{\begingroup\mn@urlcharsother \@ifnextchar [ {\mn@doi@}
  {\mn@doi@[]}}
\def\mn@doi@[#1]#2{\def\@tempa{#1}\ifx\@tempa\@empty \href
  {http://dx.doi.org/#2} {doi:#2}\else \href {http://dx.doi.org/#2} {#1}\fi
  \endgroup}
\def\mn@eprint#1#2{\mn@eprint@#1:#2::\@nil}
\def\mn@eprint@arXiv#1{\href {http://arxiv.org/abs/#1} {{\tt arXiv:#1}}}
\def\mn@eprint@dblp#1{\href {http://dblp.uni-trier.de/rec/bibtex/#1.xml}
  {dblp:#1}}
\def\mn@eprint@#1:#2:#3:#4\@nil{\def\@tempa {#1}\def\@tempb {#2}\def\@tempc
  {#3}\ifx \@tempc \@empty \let \@tempc \@tempb \let \@tempb \@tempa \fi \ifx
  \@tempb \@empty \def\@tempb {arXiv}\fi \@ifundefined
  {mn@eprint@\@tempb}{\@tempb:\@tempc}{\expandafter \expandafter \csname
  mn@eprint@\@tempb\endcsname \expandafter{\@tempc}}}

\bibitem[\protect\citeauthoryear{Bode \& Evans}{Bode \& Evans}{2008}]{bode}
Bode M.~F.,  Evans A.~e.,  1989, 2008, Classical Novae.
Cambridge University Press

\bibitem[\protect\citeauthoryear{Bonnet-Bidaud, Praderie  \&
  Whitfield}{Bonnet-Bidaud et~al.}{2019}]{bidaud}
Bonnet-Bidaud J.-M.,  Praderie F.,   Whitfield S.,  last view: 2019,
  International Dunhuang Project

\bibitem[\protect\citeauthoryear{Clark \& Stephenson}{Clark \&
  Stephenson}{1977}]{steph77}
Clark D.,  Stephenson F.,  1977, The Historical Supernovae.
Pergamon, Oxford

\bibitem[\protect\citeauthoryear{Duerbeck}{Duerbeck}{1992}]{duerbeck1992}
Duerbeck H.,  1992, MNRAS, 258, 629

\bibitem[\protect\citeauthoryear{Duerbeck}{Duerbeck}{2008}]{duerbeck}
Duerbeck H.,  2008, Novae: an historical perspective.
Cambridge University Press

\bibitem[\protect\citeauthoryear{{Gaia Collaboration}, Brown, Vallenari  \& et
  al.}{{Gaia Collaboration} et~al.}{2018}]{brown}
{Gaia Collaboration} Brown A.~G.~A.,  Vallenari A.,   et al. 2018, A+A, 616, A1

\bibitem[\protect\citeauthoryear{{G{\"o}ttgens} et~al.,}{{G{\"o}ttgens}
  et~al.}{2019}]{goettgens2019}
{G{\"o}ttgens} F.,  et~al., 2019, \aap, 626, A69

\bibitem[\protect\citeauthoryear{Hertzog}{Hertzog}{1986}]{hertzog1986}
Hertzog K.~P.,  1986, The Observatory, 106, 38

\bibitem[\protect\citeauthoryear{{Ho Peng Yoke}}{{Ho Peng Yoke}}{1962}]{ho}
{Ho Peng Yoke} 1962, Vistas in Astronomy, 5

\bibitem[\protect\citeauthoryear{Hoffmann}{Hoffmann}{2017}]{smh2017}
Hoffmann S.~M.,  2017, Hipparchs Himmelsglobus.
Springer

\bibitem[\protect\citeauthoryear{Hoffmann \& Vogt}{Hoffmann \&
  Vogt}{prep}]{hoffmannVogt2019}
Hoffmann S.~M.,  Vogt N.,  {in prep.}, Astron. Nachr.

\bibitem[\protect\citeauthoryear{Honeycutt, Robertson  \& Kafka}{Honeycutt
  et~al.}{2011}]{honeycutt}
Honeycutt R.~K.,  Robertson J.~W.,   Kafka S.,  2011, AJ, 141, 121

\bibitem[\protect\citeauthoryear{Hsi}{Hsi}{1957}]{hsi}
Hsi T.-T.,  1957, Smithonian Contributions to Astrophysics, 2

\bibitem[\protect\citeauthoryear{Jacob, Schönberner  \& Steffen}{Jacob
  et~al.}{2013}]{jacob2013}
Jacob R.,  Schönberner D.,   Steffen M.,  2013, A+A, 558, A78

\bibitem[\protect\citeauthoryear{Johansson}{Johansson}{2007}]{johansson}
Johansson G.,  2007, Nature, 448, 251

\bibitem[\protect\citeauthoryear{{Kemp} et~al.,}{{Kemp} et~al.}{2012}]{kemp}
{Kemp} J.,  et~al., 2012, in The Society for Astronomical Sciences 31st Annual
  Symposium on Telescope Science, held May 22-24, 2012 at Big Bear Lake, CA. pp
  7--15, \url {http://adsabs.harvard.edu/abs/2012SASS...31....7K}

\bibitem[\protect\citeauthoryear{{Liu Ci-Yuan}}{{Liu Ci-Yuan}}{1986}]{liu}
{Liu Ci-Yuan} 1986, Acta Astronomica Sinica, 3

\bibitem[\protect\citeauthoryear{Miszalski, Woudt, Littlefair, P., Warner  \&
  Boffin}{Miszalski et~al.}{2016}]{miszalski2016}
Miszalski B.,  Woudt P.~A.,  Littlefair P. S.,  Warner B.,   Boffin H. M.~J.,
  2016, MNRAS, 456, 633

\bibitem[\protect\citeauthoryear{Needham}{Needham}{1959}]{needham}
Needham J.,  1959, Science and Civilization in Ancient China.
~SCAC Vol. 3, Cambridge University Press

\bibitem[\protect\citeauthoryear{Nickiforov}{Nickiforov}{2010}]{nickiforov}
Nickiforov M.~G.,  2010, Bulgarian Astronomical Journal, 13, 116

\bibitem[\protect\citeauthoryear{Pankenier}{Pankenier}{2013}]{pankenier2013}
Pankenier D.,  2013, Astrology and Cosmology in Early China: Conforming Earth
  to Heaven.
Cambridge University Press

\bibitem[\protect\citeauthoryear{Patterson et~al.,}{Patterson
  et~al.}{2013}]{pat2013}
Patterson J.,  et~al., 2013, MNRAS, 434, 1902

\bibitem[\protect\citeauthoryear{Pskovskii}{Pskovskii}{1972}]{pskovskii}
Pskovskii Y.~P.,  1972, Soviet Astronomy, 16

\bibitem[\protect\citeauthoryear{Saito, Minniti, Catelan, Angeloni, Beamin,
  Palma, Gutierrez  \& Montenegro}{Saito et~al.}{2016}]{saito2016}
Saito R.,  Minniti D.,  Catelan M.,  Angeloni R.,  Beamin J.~C.,  Palma T.,
  Gutierrez L.~A.,   Montenegro K.,  2016, The Astronomer's Telegram

\bibitem[\protect\citeauthoryear{Shara et~al.,}{Shara et~al.}{2007}]{shara2007}
Shara M.~M.,  et~al., 2007, Nature, 446, 159

\bibitem[\protect\citeauthoryear{Shara, Mizusawa, Wehinger, Zurek, Martin,
  Neill, Forster  \& Seibert}{Shara et~al.}{2012}]{shara2012}
Shara M.~M.,  Mizusawa T.,  Wehinger P.,  Zurek D.,  Martin C.~D.,  Neill
  J.~D.,  Forster K.,   Seibert M.,  2012, ApJ, 756, 107

\bibitem[\protect\citeauthoryear{Shara, Drissen, Martin  \& Alarie}{Shara
  et~al.}{2016}]{shara2017_ATcnc}
Shara M.~M.,  Drissen L.,  Martin T.,   Alarie A.,  2016,
  https://arxiv.org/abs/1609.06695

\bibitem[\protect\citeauthoryear{Shara, Drissen, Martin, Alarie  \&
  Stephenson}{Shara et~al.}{2017a}]{shara2017_ATcnc_steph}
Shara M.~M.,  Drissen L.,  Martin T.,  Alarie A.,   Stephenson F.~R.,  2017a,
  MNRAS, 465, 739

\bibitem[\protect\citeauthoryear{Shara et~al.,}{Shara
  et~al.}{2017b}]{shara2017_nov1437}
Shara M.~M.,  et~al., 2017b, Nature, 548, 558

\bibitem[\protect\citeauthoryear{Stephenson}{Stephenson}{1976}]{stephenson}
Stephenson F.~R.,  1976, Quarterly Journal Royal Astronomical Society, 17, 121

\bibitem[\protect\citeauthoryear{Sun \& Kistemaker}{Sun \&
  Kistemaker}{1997}]{sunKistemaker}
Sun X.,  Kistemaker J.,  1997, The Chinese Sky during the Han.
Brill

\bibitem[\protect\citeauthoryear{{Vogt}, {Tappert}, {Puebla},
  {Fuentes-Morales}, {Ederoclite}  \& {Schmidtobreick}}{{Vogt}
  et~al.}{2018}]{vogt}
{Vogt} N.,  {Tappert} C.,  {Puebla} E.~C.,  {Fuentes-Morales} I.,  {Ederoclite}
  A.,   {Schmidtobreick} L.,  2018, MNRAS, 478, 5427

\bibitem[\protect\citeauthoryear{Vogt, Hoffmann  \& Tappert}{Vogt
  et~al.}{2019}]{vogt2019}
Vogt N.,  Hoffmann S.~M.,   Tappert C.,  2019, Astron. Nachr., pp~1--8

\bibitem[\protect\citeauthoryear{Warner}{Warner}{1995}]{warner1995}
Warner B.,  1995, Cataclysmic Variable Stars.
Cambridge Astrophysical Series, Cambridge University Press

\bibitem[\protect\citeauthoryear{Warner}{Warner}{2016}]{warner}
Warner B.,  2016, in 3rd Annual Conference on High Energy Astrophysics in
  Southern Africa - Transients. Proceedings of Science, \url {DOI:
  10.22323/1.241.0002}

\bibitem[\protect\citeauthoryear{Watson, Henden  \& Price}{Watson
  et~al.}{2006}]{watson}
Watson C.,  Henden A.~A.,   Price A.,  2006, SASS, 25, 47

\bibitem[\protect\citeauthoryear{{Xi Ze-zong} \& {Po Shu-jen}}{{Xi Ze-zong} \&
  {Po Shu-jen}}{1966}]{xi+po}
{Xi Ze-zong} {Po Shu-jen} 1966, Science, 154

\bibitem[\protect\citeauthoryear{Xu, Pankenier  \& Jiang}{Xu
  et~al.}{2000}]{xu2000}
Xu Z.,  Pankenier D.~W.,   Jiang Y.,  2000, East Asian Archaeoastronomy.
Gordon and Breach Science Publishers

\bibitem[\protect\citeauthoryear{della Valle, Gilmozzi, Bianchini  \&
  Esenoglu}{della Valle et~al.}{1997}]{valle1997}
della Valle M.,  Gilmozzi R.,  Bianchini A.,   Esenoglu H.,  1997, A+A, 325,
  1151

\makeatother
\end{thebibliography}

\appendix

%%%%%%%%%%%%%%%%%%%%%%%%%%%%%%%%%%%%%%%%%%%%%%%%%%

% Don't change these lines
\bsp	% typesetting comment
\label{lastpage}
\end{document}